\documentclass{article}

\usepackage{graphicx}
\usepackage{cite,gensymb,amssymb,url}
\usepackage[caption=false,font=scriptsize,labelfont=sf,textfont=sf]{subfig}
\usepackage{pbox}
\usepackage{threeparttable}

\begin{document}

\title{SliceType: Fast Gaze Typing\\ with a Merging Keyboard\footnote{This is a pre-print of an article published in Journal on Multimodal User Interfaces. The final authenticated version is available online at https://doi.org/10.1007/s12193-018-0285-z} \footnote{We invite the readers to test SliceType, and reach additional material from https://github.com/bbenligiray/slicetype}}

\author{Burak Benligiray \and Cihan Topal \and Cuneyt Akinlar}

\date{}

\maketitle

\begin{abstract}
Jitter is an inevitable by-product of gaze detection.
Because of this, gaze typing tends to be a slow and frustrating process.
In this paper, we propose SliceType, a soft keyboard that is optimized for gaze input.
Our main design objective is to use the screen area more efficiently by allocating a larger area to the target keys.
We achieve this by determining the keys that will not be used for the next input, and allocating their space to the adjacent keys with a merging animation.
Larger keys are faster to navigate towards, and easy to dwell on in the presence of eye tracking jitter.
As a result, the user types faster and more comfortably.
In addition, we employ a word completion scheme that complements gaze typing mechanics.
A character and a related prediction is displayed at each key.
Dwelling at a key enters the character, and double-dwelling enters the prediction.
While dwelling on a key to enter a character, the user reads the related prediction effortlessly.
The improvements provided by these features are quantified using the Fitts' law.
The performance of the proposed keyboard is compared with two other soft keyboards designed for gaze typing, Dasher and GazeTalk.
37 novice users gaze-typed a piece of text using all three keyboards.
The results of the experiment show that the proposed keyboard allows faster typing, and is more preferred by the users.
\end{abstract}

%\begin{IEEEkeywords}
%soft keyboard, on-screen keyboard, gaze typing, continuous input text entry, assistive technologies, eye tracker.
%\end{IEEEkeywords}

%~~~~~~~~~~~~~~~~~~~~~~~~~~~~~~~~~~~~~~~~~~~~~~~~~~~~~~~~~~~~~~~~~~~~~~~~~~~~~~~~~~~~~~~~~~~~~~~~~~~~~~~~~~~~~~~~~~~~~~~~~~~~~~~~
\section{Introduction}

Typing on a physical keyboard is the most typical method of text entry.
The defining characteristic of typing on a keyboard is that every character is mapped to a single key, and each keystroke results in a character being entered.
The action of typing is very simple and literal, as each atomic action results in an independent input.
This allows for easy adoption, and high performance after training.
However, as the spiritual successors of typewriters, physical keyboards have inherited many limitations.
They are designed to be used in a desktop setting, which hinders their mobile capabilities.
Additionally, they are operated by hand, which may not always be an option.

The alternative of a physical keyboard for text entry is a soft keyboard~\cite{MacKenzie:1999,MacKenzie:1999b}.
Soft keyboards can be designed to be used with discrete (e.g., buttons, blinks) or continuous input  (e.g., mouse, stylus, gaze).
They are not static like physical keyboards.
Soft keyboards can present pop-up menus~\cite{Isokoski:2004}, modify their layouts~\cite{AlFaraj:2009} and mechanics~\cite{Panwar:2012} adaptively.

The flexible nature of soft keyboards allows them to be designed to overcome limitations.
For example, mobile devices typically have small screens.
Soft keyboards for mobile devices are specifically designed to have a small footprint on the screen~\cite{MacKenzie:2002,Romano:2014}.
Head mounted displays obstruct the users' view of their environments, which disables them from getting visual feedback from a physical keyboard.
If the user cannot touch-type, a soft keyboard becomes necessary to enter text~\cite{Yu:2017}.

Users suffering from severe neural or motor disabilities (e.g., amyotrophic lateral sclerosis, locked-in syndrome) cannot use physical keyboards for text entry.
In these cases, blinks and facial gestures can be used as discrete input mechanisms~\cite{Grauman:2003,MacKenzie:2011}.
Soft keyboards designed specifically for gaze typing will be critical in improving the life quality of such users~\cite{Majaranta:2002}.
In addition, usage scenarios where non-disabled users depend solely on gaze interaction are becoming more common~\cite{Bulling:2010,Zhang:2015}.

In this paper, we propose SliceType, a soft keyboard that is optimized for gaze typing through the following features:

\begin{itemize}
	\item Based on a language model, the keys that are predicted not to be targeted next disappear, and the resulting space is allocated to the adjacent keys through a merging animation.
	Larger target keys increase text entry rate and reduce the effect of eye tracking jitter.
	\item Each key displays a character, and a related prediction based on the language model~\cite{Topal:2012}.
	Dwelling once at a key enters the character, and dwelling twice enters the prediction.
	The user reads the prediction during the first dwelling phase, which saves time and effort.
	\item The keyboard layout is designed as an inner ring composed of frequently used keys, and an outer ring composed of rarely used keys.
	In this way, travel time between frequently used keys is reduced, and the merging function is facilitated by ensuring that each frequently used key has an adjacent rarely used key.
\end{itemize}

The improvements provided by these features are quantified using the Fitts' law.
This analysis showed that the merging functionality contributes even more than word completion through predictions.
In addition, a user experiment was conducted to compare the proposed keyboard with two soft keyboards designed for gaze typing, Dasher~\cite{Ward:2000} and GazeTalk~\cite{Hansen:2001}.
The proposed keyboard was found to provide the highest text entry rate, and is preferred over the other keyboards by the users.
An interesting result of this experiment was discovering that some users' keyboard preferences were contradictory to their performances.

%~~~~~~~~~~~~~~~~~~~~~~~~~~~~~~~~~~~~~~~~~~~~~~~~~~~~~~~~~~~~~~~~~~~~~~~~~~~~~~~~~~~~~~~~~~~~~~~~~~~~~~~~~~~~~~~~~~~~~~~~~~~~~~~~
\section{Related Work}
\label{sec:related}

\begin{figure}
	\centering
	\includegraphics[width=0.8\columnwidth]{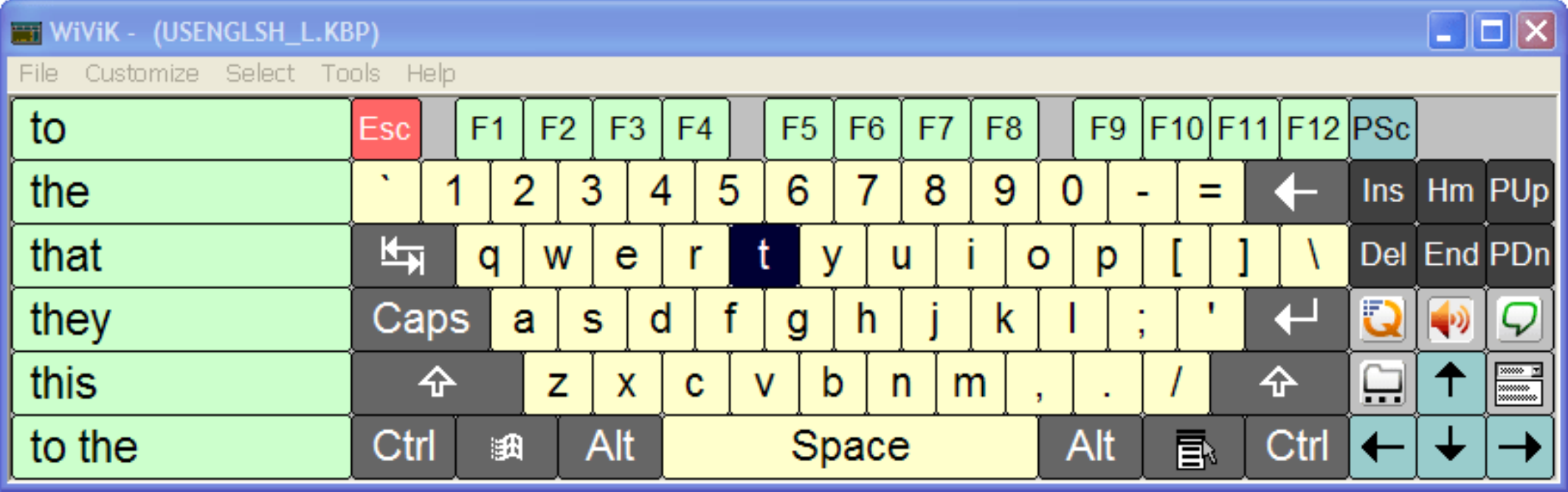}
	\caption {WiViK's user interface~\cite{Shein:1991}.
		The key being dwelt on is highlighted.}
	\label{fig:WiViK}
\end{figure}

In this section, we focus on soft keyboards that are designed to be used with a continuous input device, which can be a mouse, a touch sensitive surface, or a gaze tracker.
The input device is used either to dwell on items for selection, or to draw certain patterns through movement.
One of the earliest examples of dwelling keyboards is WiViK~\cite{Shein:1991}.
The keyboard has the traditional QWERTY layout contained in a rectangular user interface, as shown in Figure~\ref{fig:WiViK}.
To enter a character, the user moves the cursor over the key representing the character and dwells on it for a full dwell period.
WiViK predicts words that complete the current prefix, and presents them on the left.
Selecting words among these suggestions is meant to improve text entry rate.

\begin{figure}
	\centering
	\includegraphics[width=0.6\columnwidth]{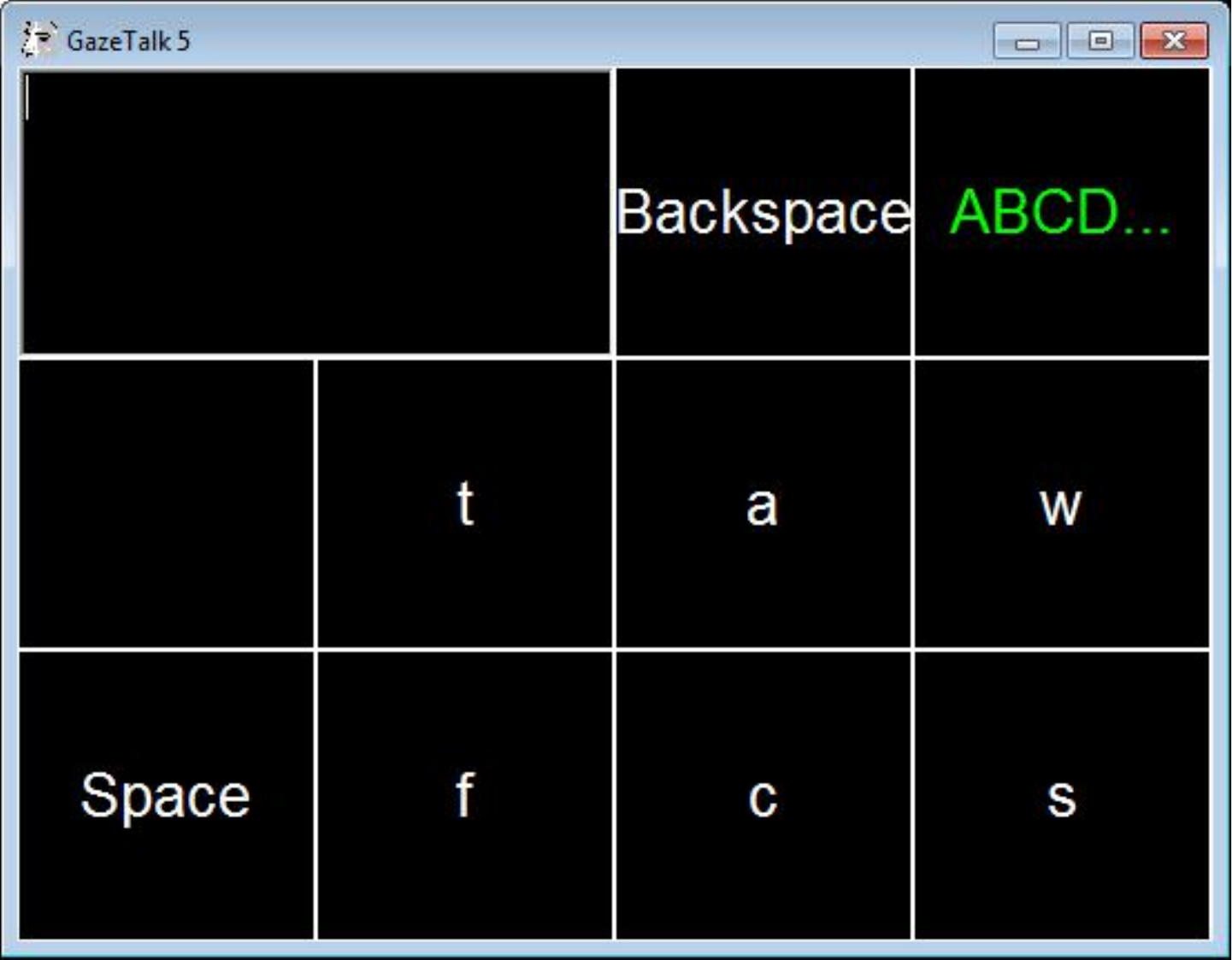}
	\caption {GazeTalk's user interface~\cite{Hansen:2001}.
		Only six characters which are most likely to be used are displayed.
		To enter another character, the user has to select the key labeled as `\texttt{ABCD\ldots}'.}
	\label{fig:GazeTalk}
\end{figure}

GazeTalk~\cite{Hansen:2001,Hansen:2003} is a dwelling soft keyboard designed specifically for gaze typing.
Its design consists of a limited number of large keys, as shown in Figure~\ref{fig:GazeTalk}.
To select a key, the user moves the cursor over it and waits for a dwell period.
A progress bar indicates the remaining dwelling time for the key to be selected.
The transcribed text is shown at the top-left key.
The key below contains a set of predictions.
Authors report a text entry rate of 4~wpm with a dwell period of 750~ms.

\begin{figure}
	\centering
	\includegraphics[width=0.6\columnwidth]{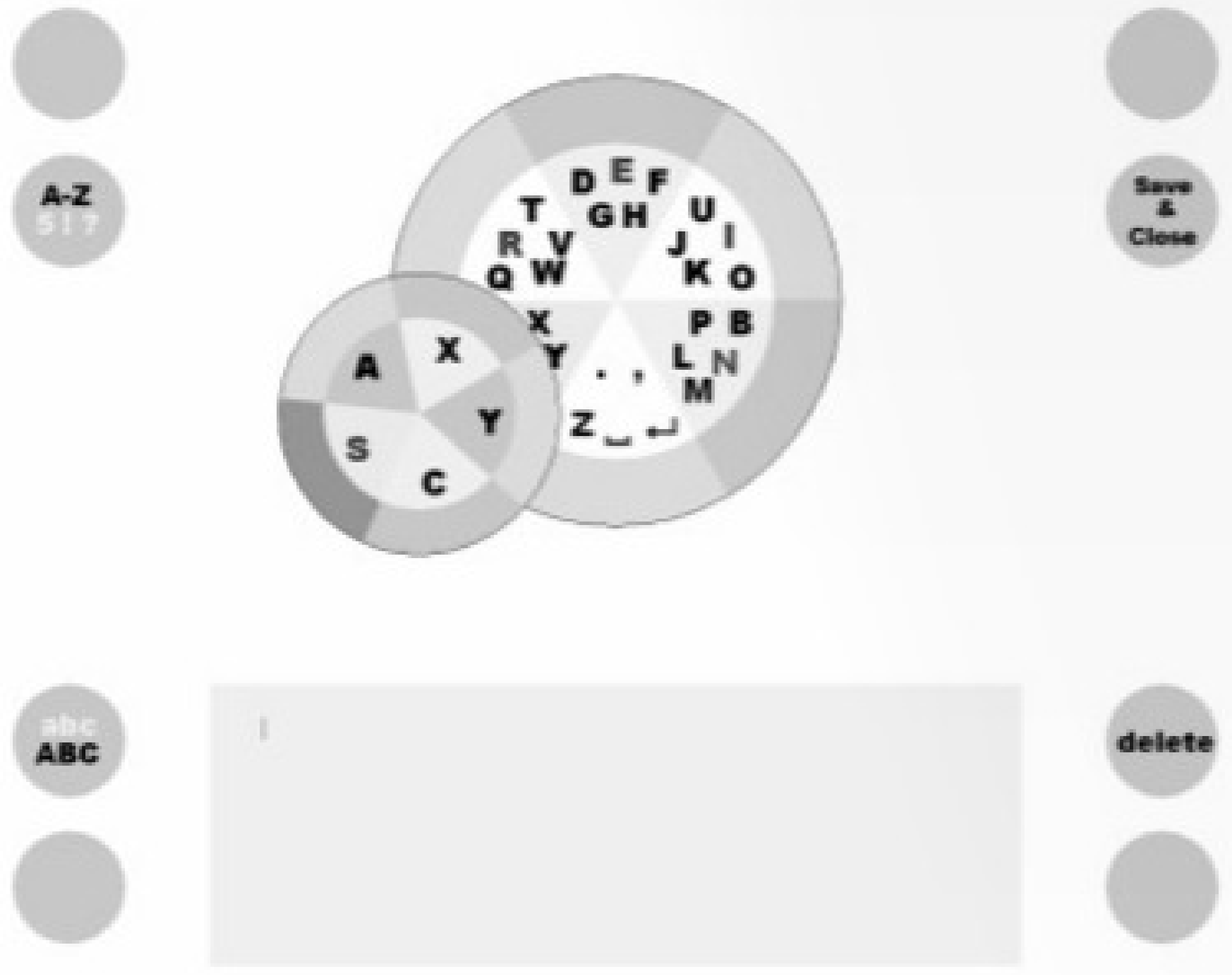}
	\caption {pEYEwrite's user interface~\cite{Huckauf:2008}.
		The circular pies in the middle are used to enter characters, which appear inside the text-box at the bottom.
		The keys on the corners provide auxiliary functionality.}
	\label{fig:pEYEWrite}
\end{figure}

pEYEwrite~\cite{Huckauf:2008} has a hierarchical interface as shown in Figure~\ref{fig:pEYEWrite}.
To enter a character, the user first dwells on the pie slice containing the desired character.
Once the dwell period is up, a circular pop-up menu appears, where a single character is displayed inside each slice.
The user dwells on the slice with the desired character for a second time to select it.
Authors report a text entry rate of 7.9~wpm with a dwell period of 400~ms.

For dwelling soft keyboards, reducing the dwell period allows for a higher text entry rate, but it also increases typing errors.
Most systems use a fixed time interval in the range of 500--1000~ms.
Alternative approaches are to adjust the dwelling period adaptive to user performance~\cite{Panwar:2012}, or to use different dwell periods for different keys~\cite{Mott:2017}.

\begin{figure}
	\centering
	\includegraphics[width=0.8\columnwidth]{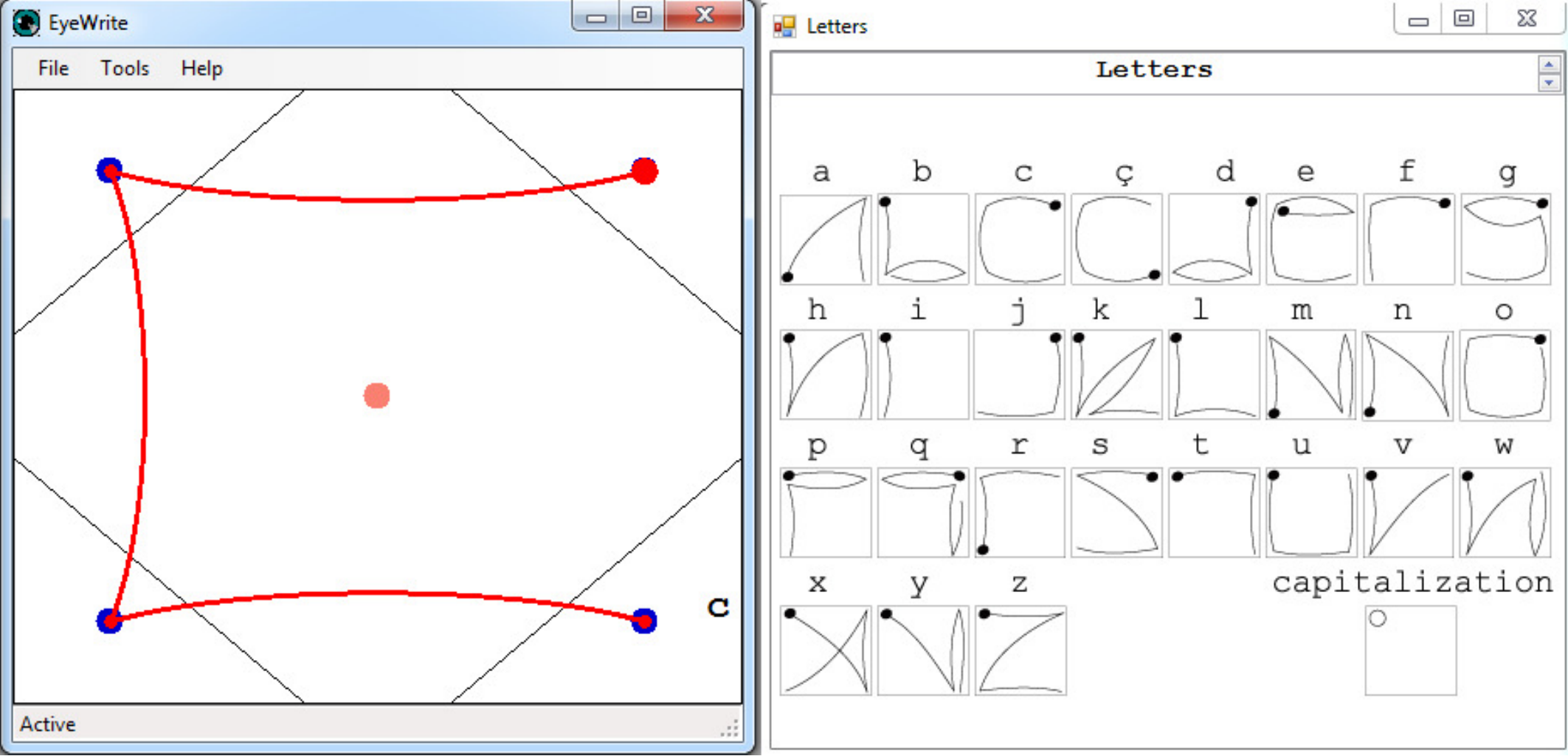}
	\caption {EyeWrite's user interface and the patterns representing each character~\cite{Wobbrock:2008}.
		In the figure, the letter `\texttt{c}' is being entered.}
	\label{fig:EyeWrite}
\end{figure}

EyeWrite~\cite{Wobbrock:2008} is an alternative system, where entering a character involves drawing a certain pattern by moving the cursor, rather than dwelling on a key.
Patterns resemble pen strokes for writing letters, and are drawn by moving the cursor from one corner of a rectangular window to another, as shown in Figure~\ref{fig:EyeWrite}.
Authors report a text entry throughput of about 5~wpm.
In addition to EyeWrite, Minimum Device Independent Text Input Method (MDTIM)~\cite{Isokoski:2000} and Eye-S~\cite{Porta:2008} also employ this style of ``eye graffiti'' communication.
In these systems, letters are created by a sequence of fixations on regions called hot-spots.
These regions can be made invisible and do not interfere with other applications, meaning that the entire screen area can be utilized by other applications.

EyeSwipe~\cite{Kurauchi:2016} has a QWERTY layout, but keys are not selected by dwelling.
Instead, the user glances through the keys in sequence, similar to the swiping motion used in soft keyboards for mobile devices.
Memorization of the layout is critical for this system, as the user should not scan the keys to search for a character in the middle of typing word.

\begin{figure}
	\centering
	\includegraphics[width=0.8\columnwidth]{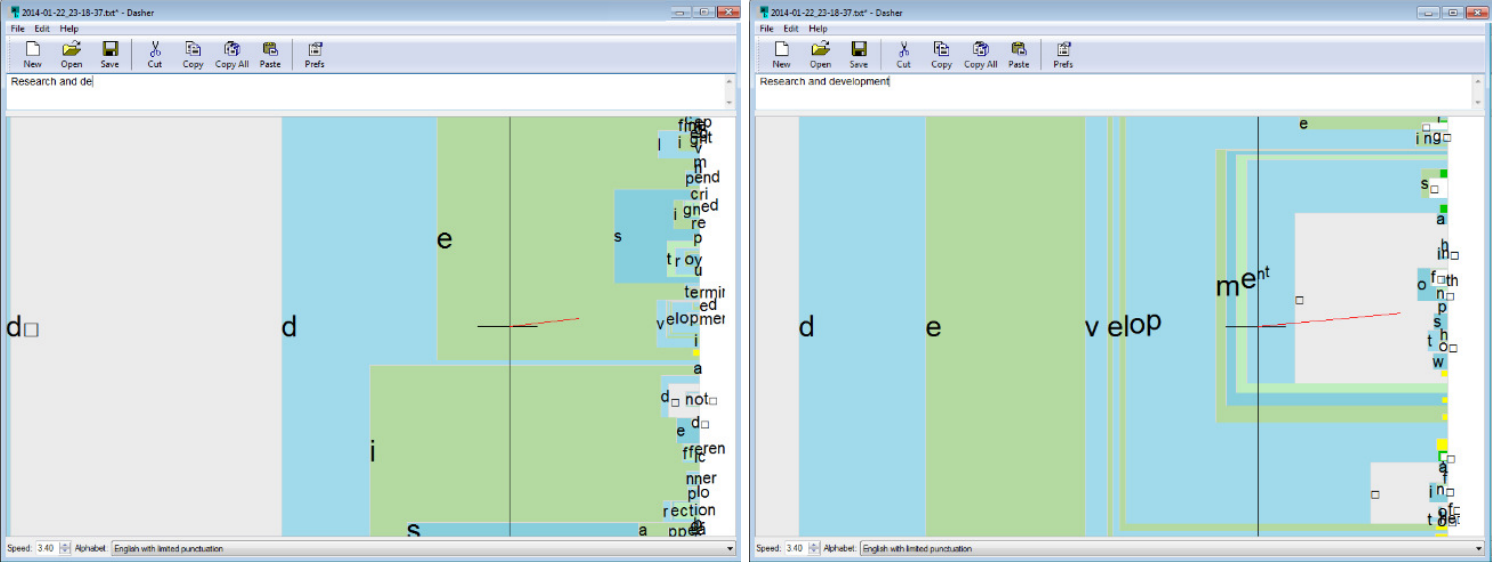}
	\caption {Dasher's user interface~\cite{Ward:2000}.
		Text is entered by moving the cursor towards the desired characters.
		In the figure, the word ``\texttt{development}" is being written.}
	\label{fig:Dasher}
\end{figure}

Dasher~\cite{Ward:2000} has a radically different design that is used by a continuous gesture.
Characters appear on the right hand-side of the interface as shown in Figure~\ref{fig:Dasher}.
To select a character, the cursor is moved towards it.
This causes the area representing the target character to become larger, which is called dynamic zooming.
When the character crosses over the vertical line, it is entered.
While approaching the target character, the characters that may follow it appear inside its area.
To undo an entry, the cursor must be moved to the left hand-side of the vertical line.
This causes the entered characters to move back to the right hand-side of the vertical line and be removed.
Authors report a text entry rate of about~20 wpm after one hour of practice.

\begin{figure}
	\centering
	\subfloat[]
	{
		\includegraphics[width=0.34\columnwidth]{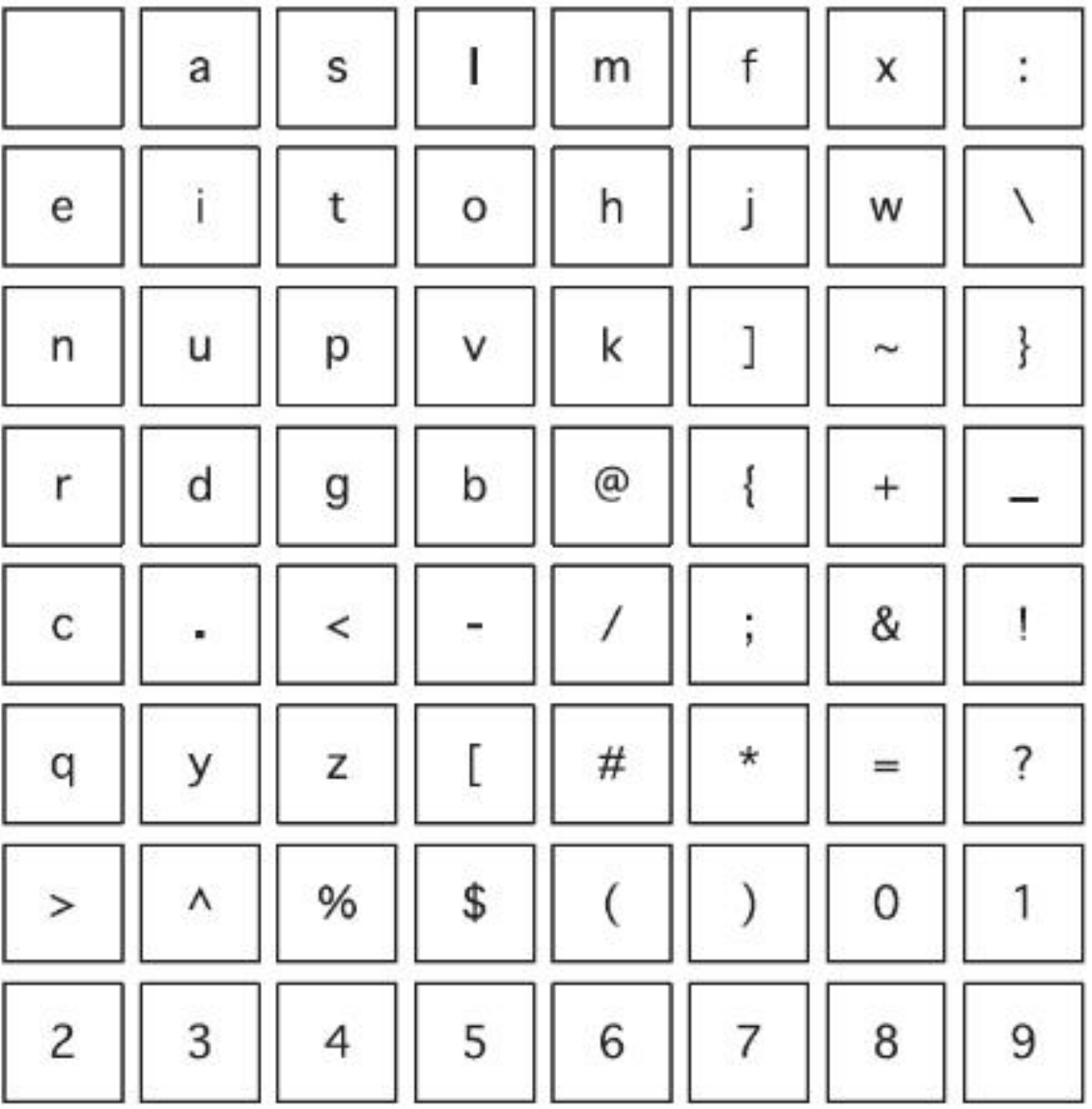}
	}
	\subfloat[]
	{
		\includegraphics[width=0.44\columnwidth]{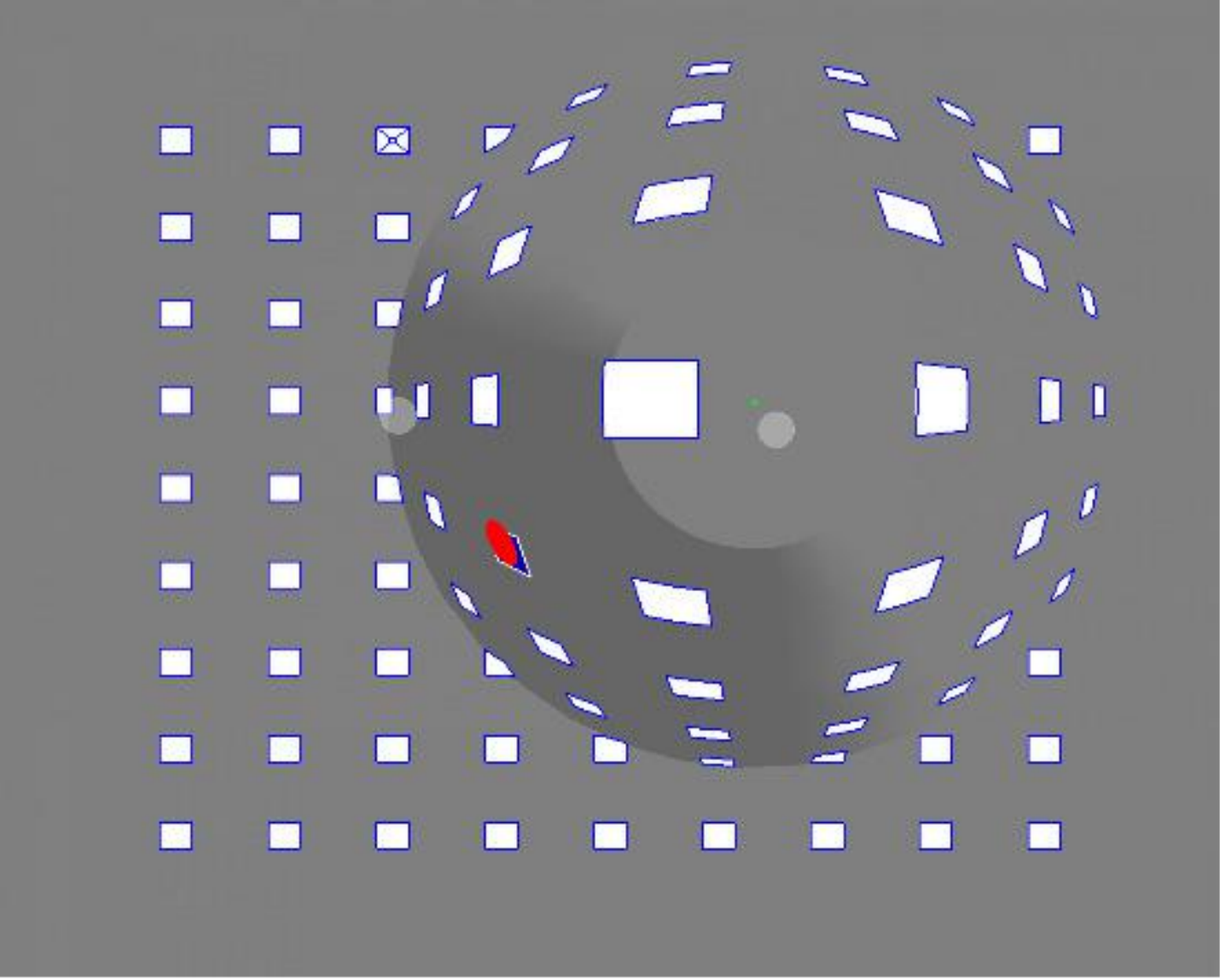}
	}
	\caption {(a)~Zooming user interface to make easy and correct selections~\cite{Ashmore:2005},
		(b)~A keyboard layout to minimize the cursor movement to enter a book chapter~\cite{Francis:2011}.}
	\label{fig:ZoomingInterface}
\end{figure}

Zooming is another interesting concept in which the region having the current user focus is made bigger for easy and accurate selection~\cite{Bates:2002,Ashmore:2005}.
Figure~\ref{fig:ZoomingInterface}~(a) illustrates this idea in action.
A study on how auditory and visual feedback affects gaze typing is presented in~\cite{Majaranta:2003}.
Authors state that proper feedback by the system influences both the text entry rate and error rate.

There is also a good body of research on the optimal layout and size of the keys on the soft keyboard interface~\cite{Venkatagiri:1999,Zhai:2000,Zhai:2002,Bates:2002,Ashmore:2005}.
For example, Figure~\ref{fig:ZoomingInterface}~(b) shows a keyboard layout optimized to minimize the text entry time for a specific book chapter.
Word or phrase completion or prediction is widely employed by many soft keyboards\cite{Foster:2002,Grabski:2004,Bickel:2005,MacKenzie:2008,Sharma:2010}.
The idea with word completion is to look at the previously entered text and make suggestions based on a language model.
The goal is to allow the user enter the complete word without entering the remaining letters, thus increasing text entry rate.
Sharma  et al. investigate the effect of prediction list orientation, position, and the number of predictions to be displayed~\cite{Sharma:2010}.

Let us investigate the design aspects of the aforementioned soft keyboards, and if they are suitable for gaze typing.
The most typical characteristic of gaze input is its inaccuracy, which can manifest as a constant bias or random jitter.
This can be combated by ensuring that the interface components are sufficiently large.
Since large components are desired and the total area is limited, efficiency is a key aspect of good soft keyboard design.
From this standpoint, no area on the interface should remain unutilized, which is not universally adhered to (e.g., pEYEwrite).
In relation to this, allocating a separate area for predictions as WiViK and GazeTalk is a common practice that wastes area and causes constant diversion of attention.
Both Dasher and the proposed design circumvent the need for a static prediction component that takes up space.

If the input resolution is low, hierarchical selection is a must.
However, having to do multiple actions per entry puts a hard cap on the potential entry rate.
Gaze input is noisy, rather than low-resolution.
Thus, trying to compensate for this noise by using larger elements is a more suitable alternative for gaze typing.
Therefore, pEYEwrite, GazeTalk and other soft keyboards that utilize hierarchical selection are not optimal for gaze typing, especially considering that gaze tracking devices grow in accuracy by the day.

The final design aspect that we would like to touch on is the balance between static and dynamic layouts.
Hard keyboards are fully-static, which allows the experienced user to touch type with great speed and comfort.
On the other hand, soft keyboard layouts can change dynamically, which opens up a path for interesting designs.
For optimal performance and comfort, a balance needs to be struck.
We believe that the most important rule regarding this trade-off is that the positions of keys should not change dramatically.
A completely dynamic layout as in GazeTalk and Dasher requires constant cognitive effort from the user, as the location of each key is unpredictable.
It is to be expected that such an effort will have negative consequences on typing performance and comfort.

%~~~~~~~~~~~~~~~~~~~~~~~~~~~~~~~~~~~~~~~~~~~~~~~~~~~~~~~~~~~~~~~~~~~~~~~~~~~~~~~~~~~~~~~~~~~~~~~~~~~~~~~~~~~~~~~~~~~~~~~~~~~~~~~~
\section{Proposed Soft Keyboard Design}
\label{sec:slicetype}

In this section, the design of the proposed dwelling soft keyboard, SliceType, will be discussed.
Some of these discussions apply for any kind of continuous input, while others are gaze typing-specific.
With a conventional continuous input device, the user locates the item to be selected, drags the cursor upon the item, and selects it.
In gaze typing, where the selection cursor moves with gaze, the location and dragging occur concurrently.
As soon as the user sees the item to be selected, the cursor will be on the said item.
Then, the item is selected by dwelling.
The user has to keep looking at the item to dwell, thus the dwelling period cannot be used to search for the next target, or check a dedicated prediction area.
For these reasons, dwelling keyboards for gaze typing require special design considerations.

The implementation of an efficient word prediction proposal method is discussed in Section~\ref{sec:wordpred}.
In Section~\ref{sec:keylayout}, we present the layout of the proposed keyboard.
The mechanics of key merging are explained in detail in Section~\ref{sec:keymerging}.
A brief usage example is presented in Section~\ref{sec:usage}.

\subsection{Word Prediction}
\label{sec:wordpred}

Word prediction is a common tool used to speed up typing for soft keyboards.
Using the previously entered text and a language model, the current word to be entered is predicted and proposed to the user for faster entry.
Word prediction is generally considered to be limited to the generation of the prediction, using the recently entered characters and statistical data.
However, in the soft keyboard interface context, the method of communicating the predictions to the user is a problem by itself.
The generation of the prediction is more in the scope of natural language processing, rather than human--computer interaction.
Therefore, we will briefly explain the simple prediction generation scheme used in SliceType, then move on to how the predictions are presented to the user.

\subsubsection{The Prediction Engine}
\label{sec:predengine}

A bigram corpus that consists of word pairs and a unigram corpus that consists of single words are maintained. 
The last word entered and the current word prefix are used to predict the word that the user is currently entering. 
If there are multiple candidates for prediction, the word that belongs to the more frequently used bigram pattern is preferred.
If there are no suitable bigrams, unigram prediction is used as backup.

If the corpora would have been kept as a list, finding a prediction would become more complex as the sizes of the corpora grow. 
Instead, the unigram and bigram corpora are kept as individual tries. 
These tries can be updated by adding words, or arranging words based on usage frequency, which will result in a prediction engine that dynamically adapts to the user.
Refer to~\cite{Manning:1999} for more information on n-gram language models.

\subsubsection{Suggesting the Predictions}
\label{sec:relay}

The predictions are generally displayed in a dedicated area of the interface (see Figure~\ref{fig:WiViK}).
This is because they are seen as an addition to improve performance, and not as an essential part of the soft keyboard.
However, this separation has many disadvantages for gaze typing:

\begin{itemize}
	\item While gaze typing, every time the user glances at the predictions, they are also dragging the cursor over.
	This reflects as a direct Fitts' law cost~\cite{Fitts:1954}.
	\item Since the area dedicated for the predictions is limited, not many predictions can be displayed.
	It is less likely to be able to propose correct predictions with fewer guesses.
	\item The user has to read a prediction list until they encounter the desired word.
	If the desired word is not among the predictions, the user will have to read all of the predictions, which has the highest time cost.
	\item It is easy for the user to give up on the predictions entirely, e.g., when many false predictions are proposed consecutively.
	This also presents itself in novice users, where the user enters a state of tunnel vision and simply ignores the predictions to avoid complexity, which delays acquiring the skills needed to use the keyboard efficiently.
\end{itemize}

We quantified the negative effects of presenting the predictions in a dedicated area, and not using the predictions in Section~\ref{sec:fitts}.

The conventional prediction engines use the previously entered text to find the most probable continuations.
Instead, we propose to add each of the letters to the previously entered text and make a respective prediction, as first used in~\cite{Topal:2012}.
Since there are 26 letters in the English alphabet, this will result in 26 different predictions.
Each of these predictions will be presented in the respective key.
While the user is dwelling on a key to select it, they will read the associated prediction effortlessly.
If they wish to use the prediction, they keep on dwelling for an additional dwell period.
Otherwise, they move on to the next character.
The proposed prediction proposal method improves upon the conventional methods regarding all of the problems stated above.
A disadvantage is that every key must be large enough to encase a legible word, which is a problem for keyboards with small keys~\cite{DiazTula:2016}.
In this study, we have partly overcome this problem by enlarging the target keys through merging.

\subsection{Layout}
\label{sec:keylayout}

\begin{figure}
	\centering
	\includegraphics[width=0.6\columnwidth]{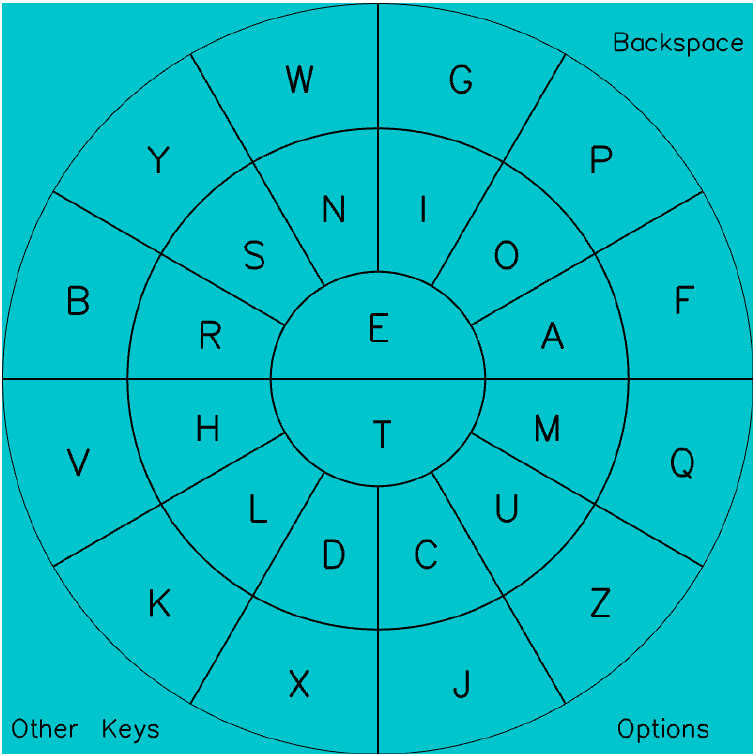}
	\caption {SliceType interface and the layout of the characters.}
	\label{fig:SliceTypeGUI}
\end{figure}

A dwelling keyboard can either be designed to display all characters at once, or group characters under categories, making them reachable by a hierarchical selection mechanism.
The need for an additional selection event for each character slows down typing, but only requires a small number of keys to be displayed at once.
Less and larger keys can be selected easily with lower resolution input devices.
On the other hand, an accurate input device is needed to use a keyboard that displays all characters at once.
Considering that eye tracking technology has reached some level of maturity, and will only improve in the future, we opted for displaying all characters for SliceType (see Figure~\ref{fig:SliceTypeGUI}).

The keyboard is shaped as a circle encased in a square. The aspect ratio is set to be 1:1, which provides compactness to the keyboard, while minimizing the distance among the characters.
The circular keyboard area contains the characters, while the corners are used for additional functionality.
The circular area has an innermost circle enclosed by two outer rings.
The innermost circle is divided in half, and the outer rings are divided radially into 24 keys, resulting in a key for each letter of the English alphabet.

The typical design objective in the character layout is to minimize the total distance traveled during typing.
An obvious solution to this problem is to assign the more frequently visited characters towards the center of the interface.
To achieve this, we calculated the frequency statistics of the letters in our English corpus.
We placed the two most frequently used letters, `\texttt{e}' and `\texttt{t}' in the two halves of the innermost circle.
The remaining 12 frequently used letters are placed in the inner ring, and the 12 rarely used letters are placed in the outer ring, as shown in Figure~\ref{fig:SliceTypeGUI}.

Reducing the cursor movement is not our only goal, which is why we avoided the common strategy of placing letters that are likely to follow each other adjacently.
We also want to facilitate key merging, and placing letters that are likely to follow each other actively prevents this.
Instead, the layout described above ensures that each letter has at least one rarely used letter adjacent to it.

\subsection{Key Merging}
\label{sec:keymerging}

The Fitts' law indicates that user movement to a target becomes faster as the target width increases~\cite{Fitts:1954}.
Furthermore, gaze tracking systems introduce some error to the system, which manifests itself as a jittering cursor.
While dwelling on the target, it is undesirable for the cursor to leave the target due to noise, which will reset the dwell timer and waste time.
Then, increasing the size of the target keys will improve speed and reduce the probability of errors.

It can be agreed on that larger keys are easier to select.
However, the total area of the graphical interface is limited.
In this case, the only way of allocating more area to a key is removing area from another key.
Allocating a smaller area to the keys that are less likely to be selected~\cite{AlFaraj:2009} may result in a frustrating experience when they are to be selected.
Instead, a more drastic approach is removing the keys that are less likely to be selected.
Doing so will result in more area to be allocated to the keys that are more likely to be chosen.
This removal of a key and using its space for an adjacent key is defined to be key merging, and is employed in SliceType.

\begin{figure}
	\centering
	\includegraphics[width=0.6\columnwidth]{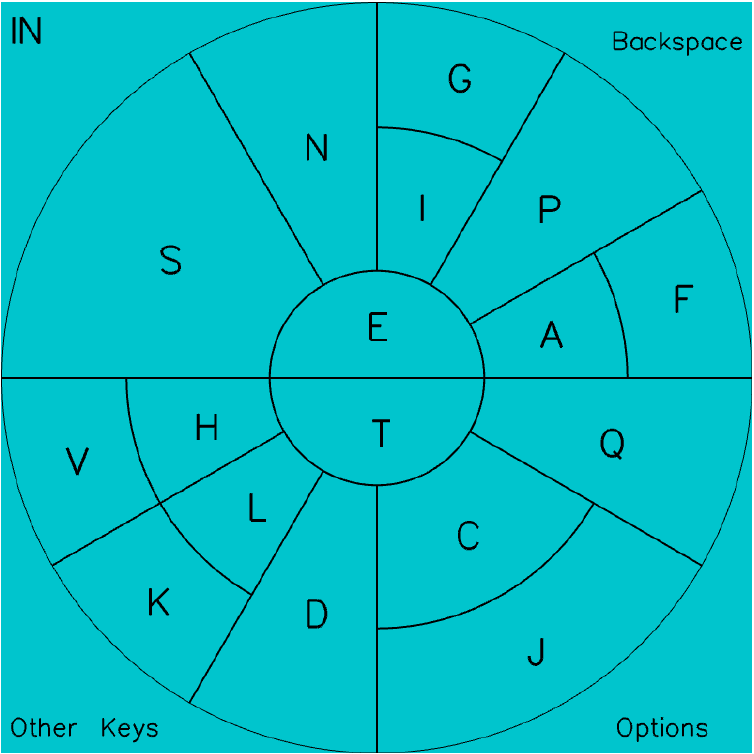}
	\caption {Merging of the keys based on the current word prefix ``\texttt{in-}".}
	\label{fig:SliceTypeGUI2}
\end{figure}

To decide on which keys are to be eliminated, we use the word predictions.
See Figure~\ref{fig:SliceTypeGUI2} for an example.
The text entered so far, ``\texttt{in-}", appears on the top-left corner.
The prediction engine does not return a result for the key `\texttt{y}', because there are no words that start with ``\texttt{iny-}".
Thus, the key is removed from the interface, and its space is merged into its neighbor, key `\texttt{s}'.
Similarly, key `\texttt{w}' is removed and its space is merged into key `\texttt{n}'.
Notice that if letters that frequently follow each other were specifically placed next to each other on the interface, such merging would have been less frequent.

The proposed key merging method assumes that the user enters a word that is present in the corpus.
In the case that the word to be entered is not in the corpus, the user will be aware when the following key to be entered merges to another key, hence does not appear on the screen.
Then, the user can undo the typing actions for the last word and enable a non-merging mode of the keyboard that allows typing words that are not present in the corpus.
This mode also adds this word to the corpus, so that the user can enter the word in the regular operation mode next time.

\begin{figure}
	\centering
	\includegraphics[width=0.6\columnwidth]{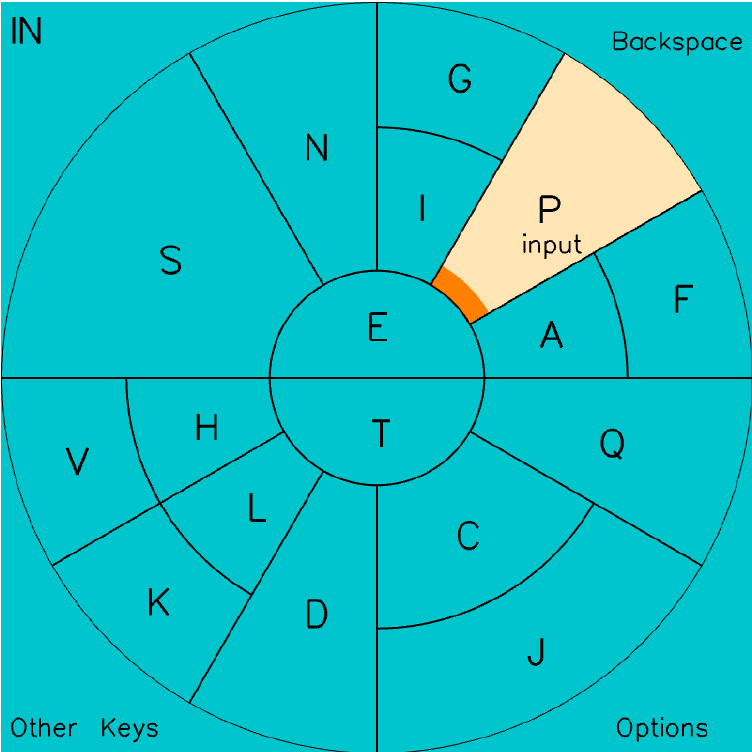}
	\caption {Using the current prefix ``\texttt{in}" and the next letter `\texttt{p}', SliceType suggests the word ``\texttt{input}", which appears inside the pie slice representing letter `\texttt{p}'.}
	\label{fig:SliceTypeGUI3}
\end{figure}

See Figure~\ref{fig:SliceTypeGUI3} for the prediction proposal mechanism.
As seen on the upper-left corner, the user has typed ``\texttt{in-}".
The user intends to select the letter `\texttt{p}' by hovering over it.
Now, not only SliceType highlights the slice representing the currently selected letter `\texttt{p}' in orange, but it also displays the word ``\texttt{input}" (generated using the current word prefix ``\texttt{inp-}").
Displaying the suggested word in the same slice as the current letter enables the user to read the suggestion without directing their gaze to a different location.
Notice that SliceType only displays the prediction within the key that is being dwelt on.
Displaying all predictions at all times would have cluttered the interface, which would hinder the user's peripheral vision from locating the next target.

\subsection{Usage}
\label{sec:usage}

This section is intended to further clarify the concepts introduced in Section~\ref{sec:slicetype}, and serve as an instruction guide for testing the proposed keyboard.
SliceType uses a direct selection system, where letter selection is performed by dwelling inside the key representing the letter.
The default dwell period is 1000 ms, which is intended to be used by novice users.
The user is expected to decrease this parameter manually as they gain confidence.

\begin{figure}
	\def\mywidth{0.48\columnwidth}
	\centering
	\subfloat[]
	{
		\includegraphics[width=\mywidth]{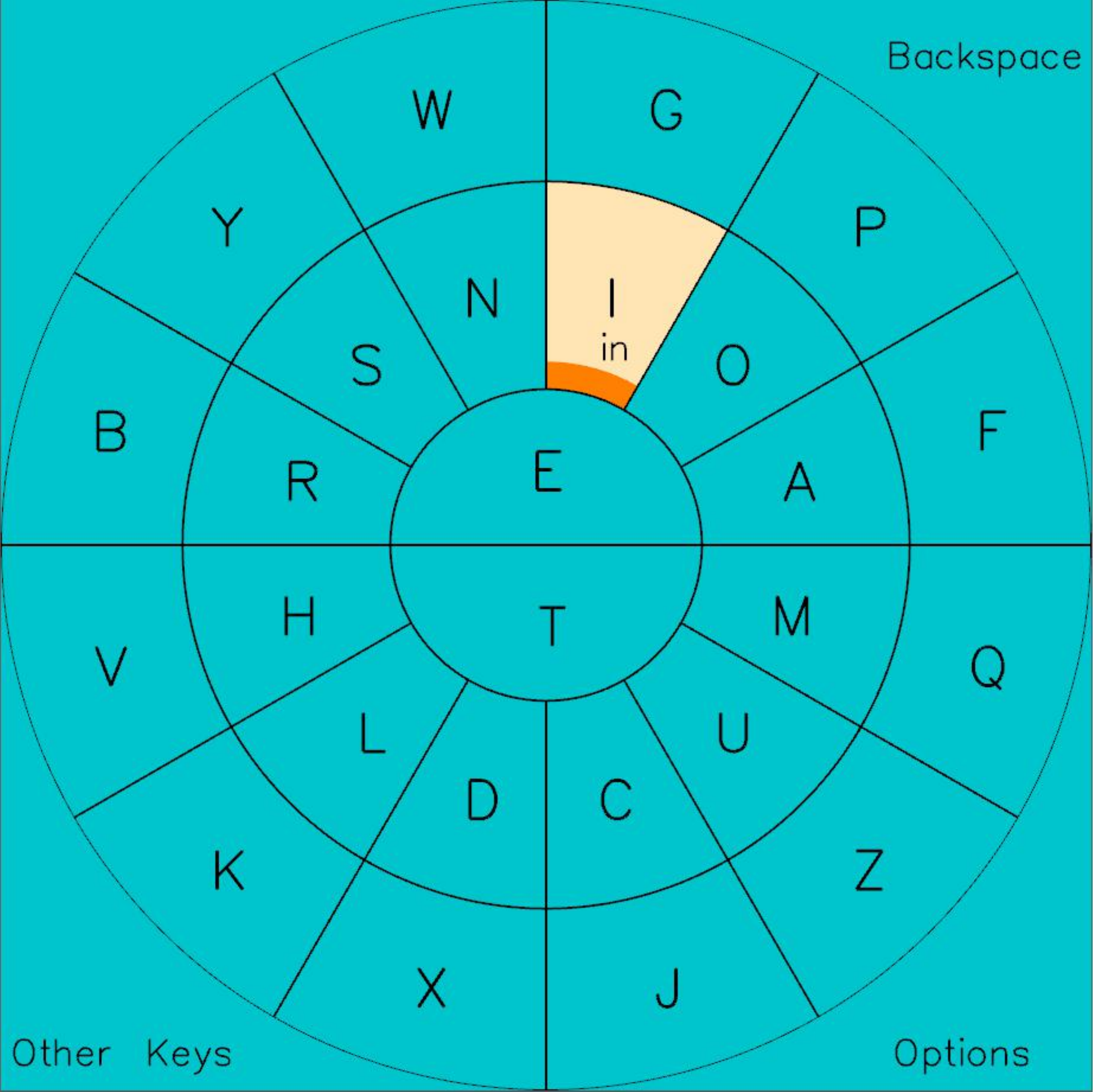}
	}
	\subfloat[]
	{
		\includegraphics[width=\mywidth]{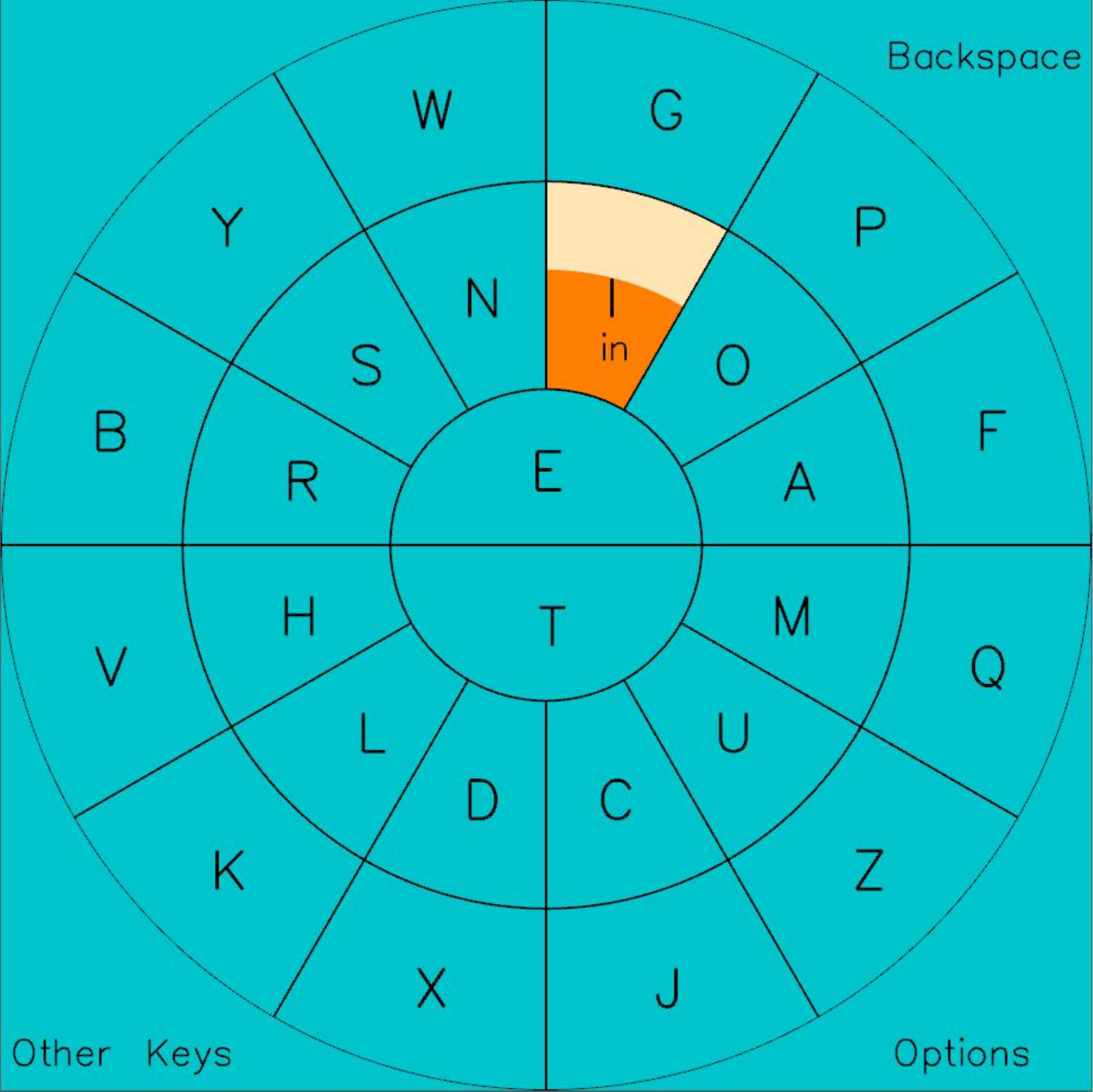}
	}
	\vspace{-\baselineskip}
	
	\subfloat[]
	{
		\includegraphics[width=\mywidth]{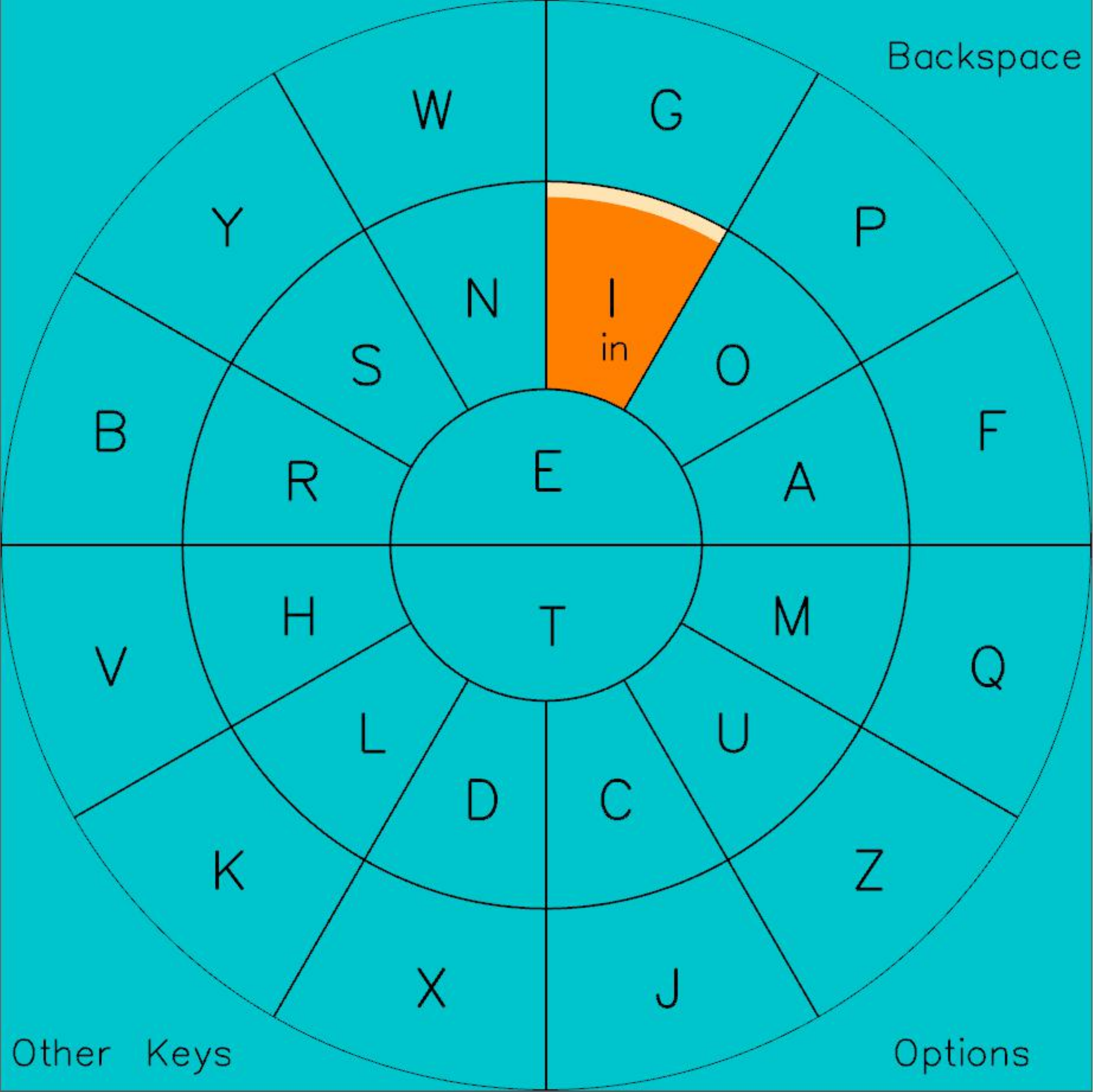}
	}
	\subfloat[]
	{
		\includegraphics[width=\mywidth]{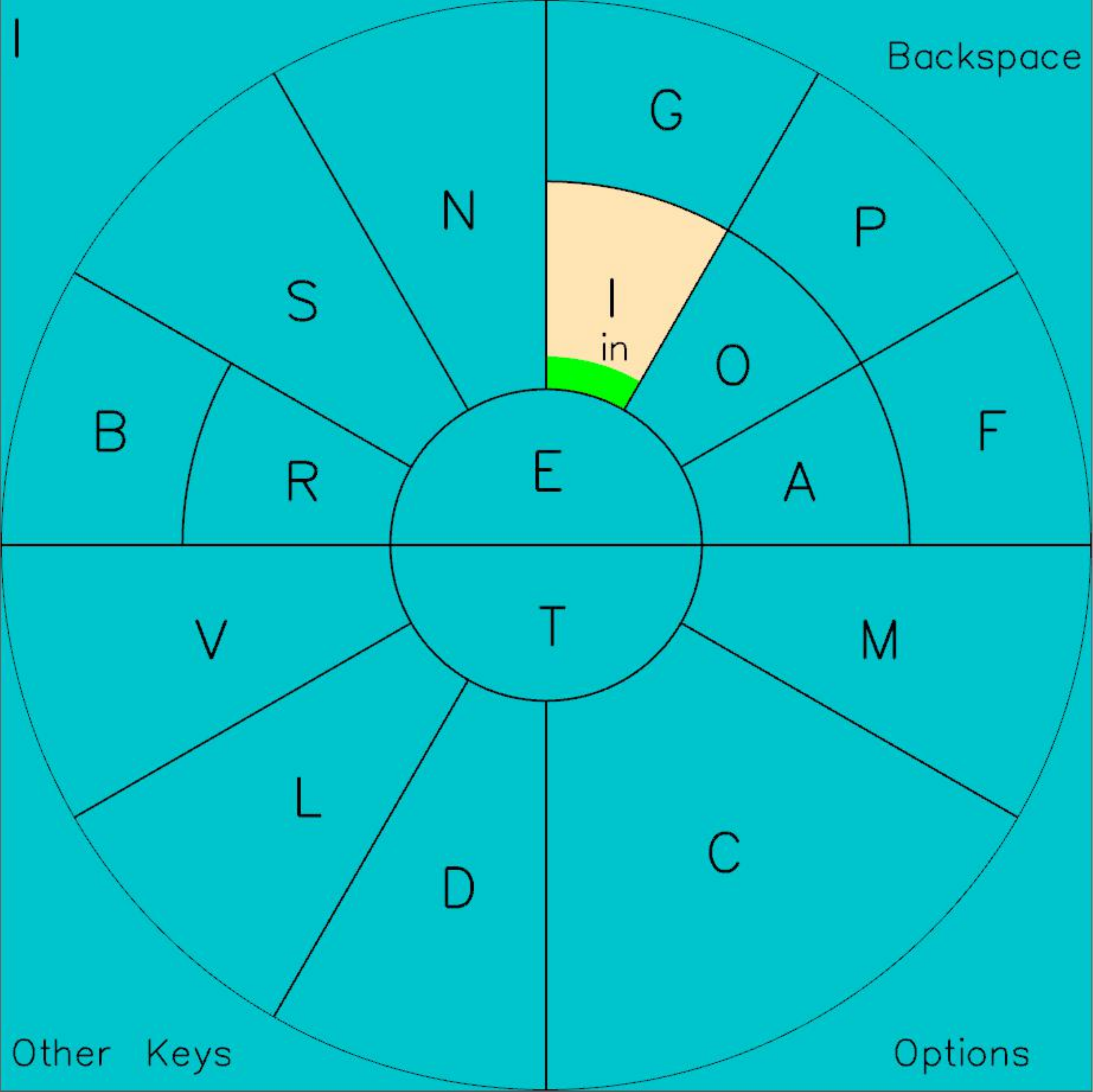}
	}
	\vspace{-\baselineskip}
	
	\subfloat[]
	{
		\includegraphics[width=\mywidth]{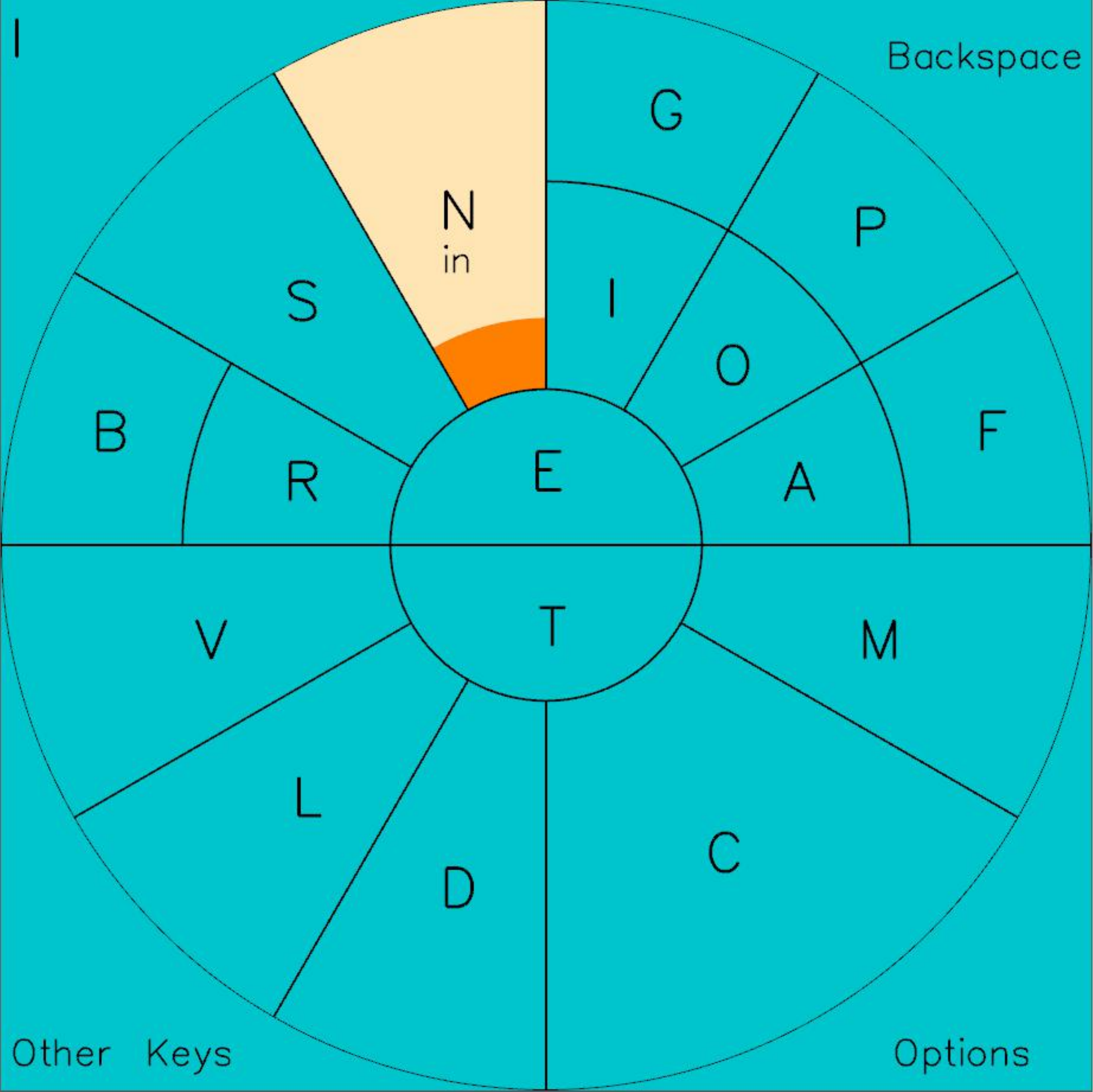}
	}
	\subfloat[]
	{
		\includegraphics[width=\mywidth]{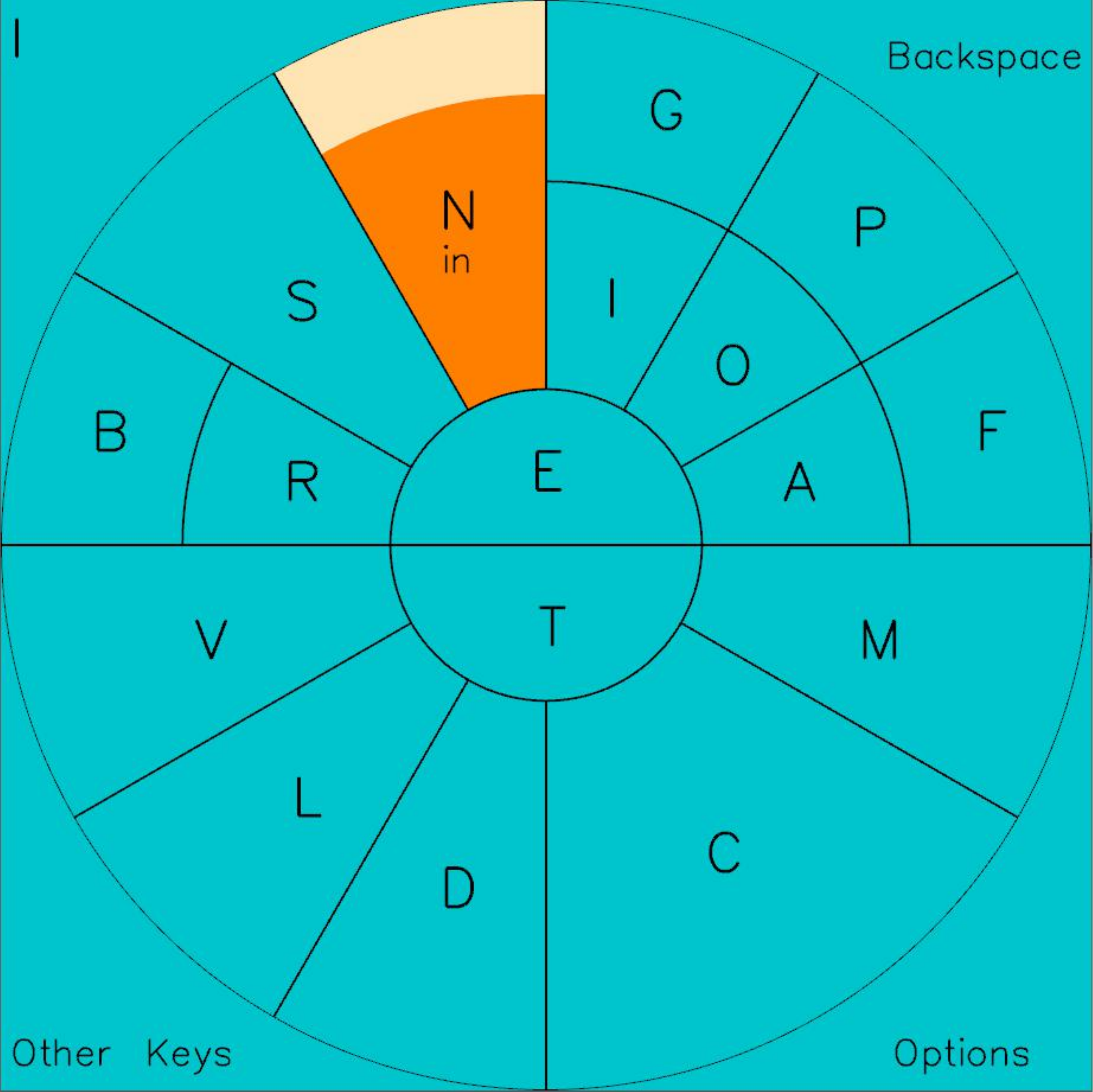}
	}
	\caption {(a) The user has just started dwelling upon the letter `\texttt{i}', which is highlighted in light orange. 
		The prediction ``\texttt{in}" is proposed. 
		(b) About 50\% of the dwell period is up.
		(c) About 90\% of the dwell period is up.
		(d) The entire dwell period is up.
		The letter `\texttt{i}' is selected and appears on the top-left corner.
		(e) The user moves on to the letter `\texttt{n}'. Note that the same prediction can be proposed again.
		(f) The user keeps dwelling on `\texttt{n}'.}
	\label{fig:SliceTypeUsage1}
\end{figure}

\begin{figure}
	\centering
	\def\mywidth{0.48\columnwidth}
	\subfloat[]
	{
		\includegraphics[width=\mywidth]{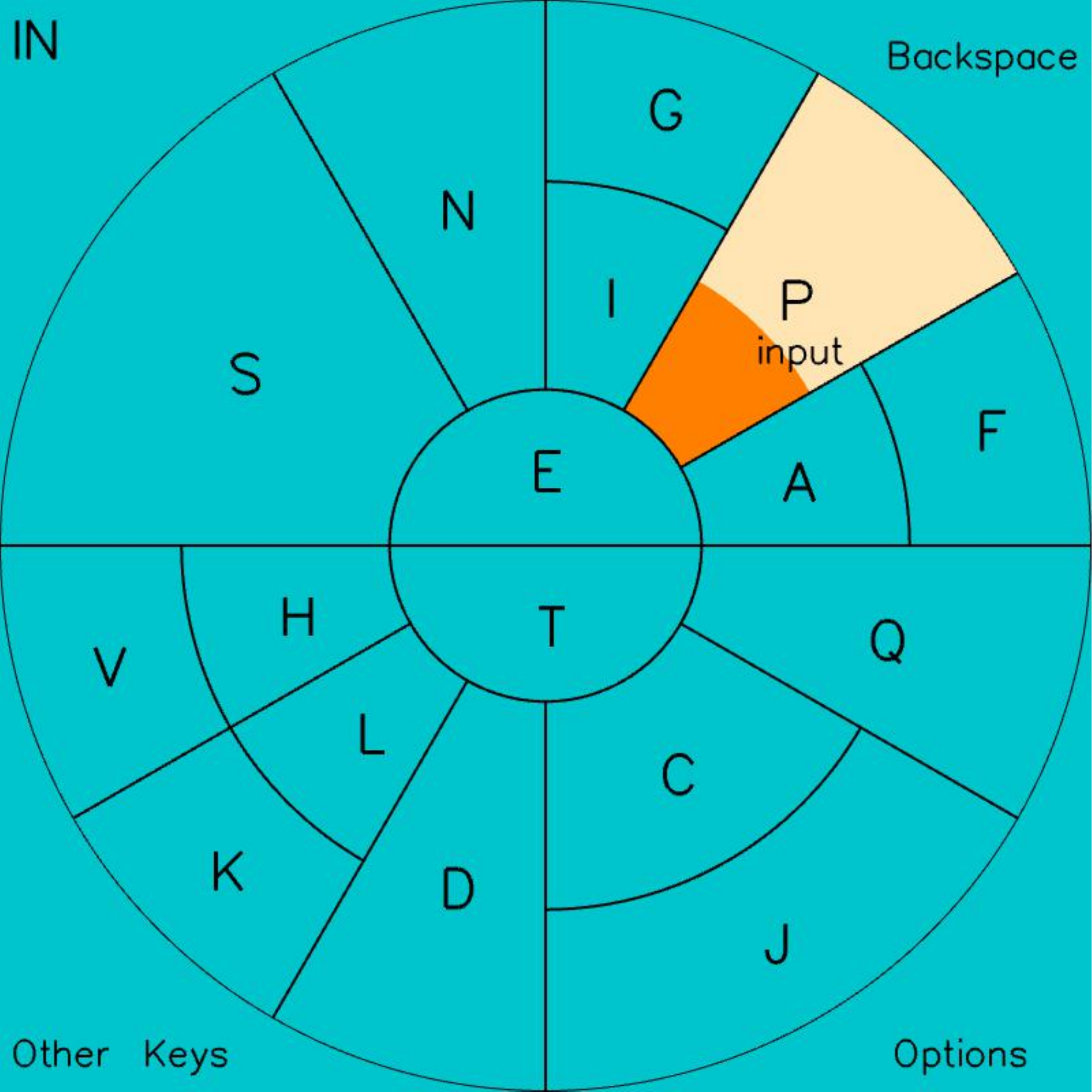}
	}
	\subfloat[]
	{
		\includegraphics[width=\mywidth]{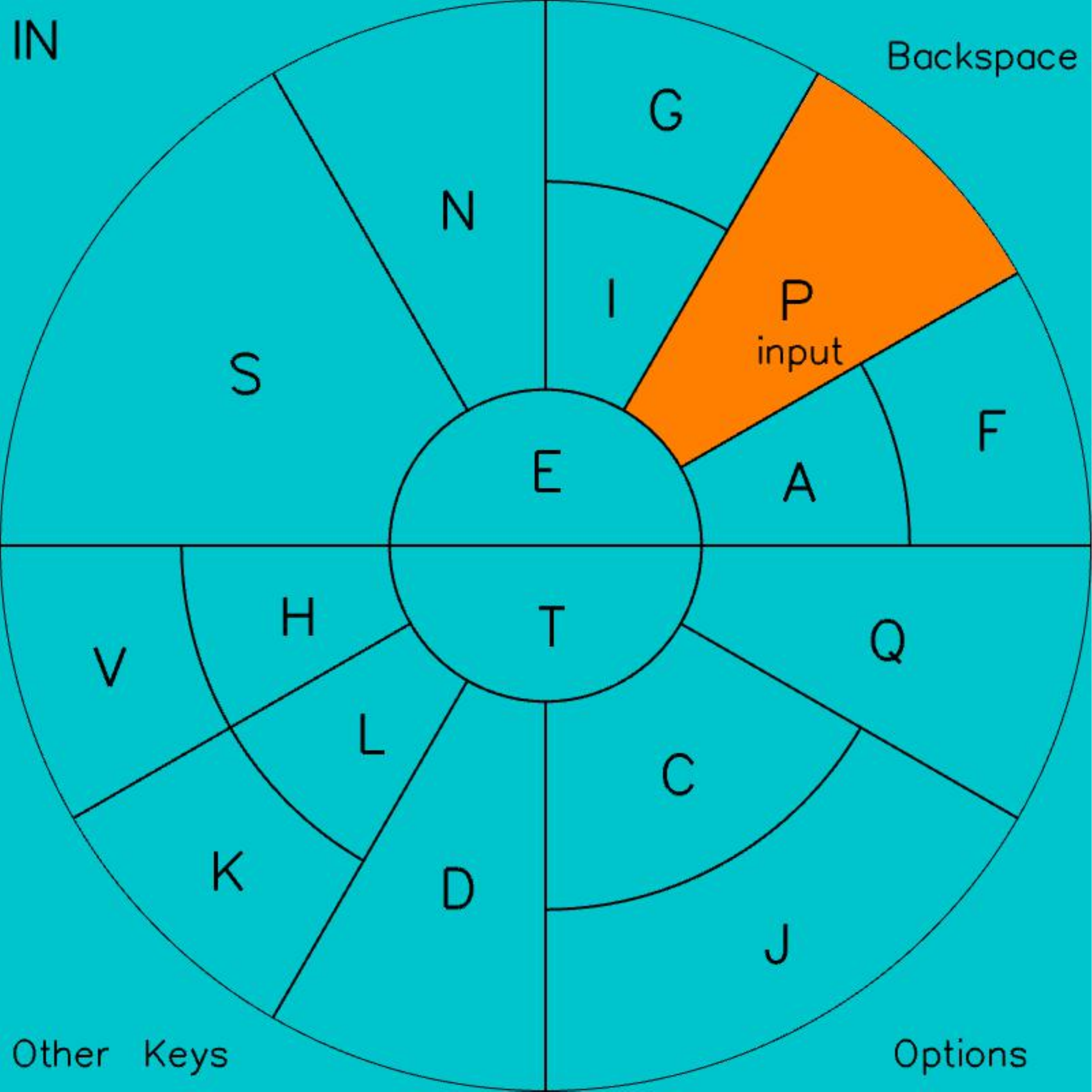}
	}
	\vspace{-\baselineskip}
	
	\subfloat[]
	{
		\includegraphics[width=\mywidth]{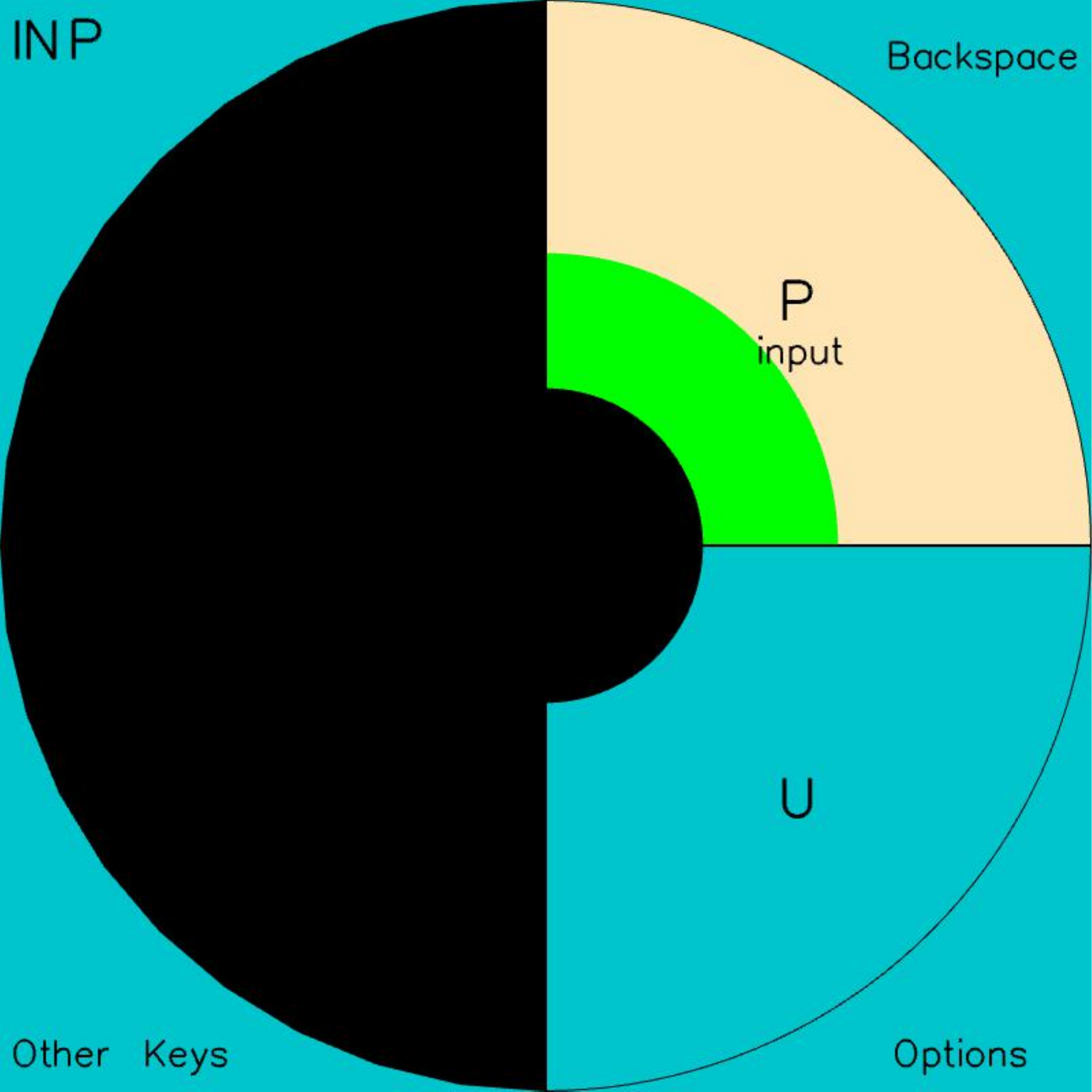}
	}
	\subfloat[]
	{
		\includegraphics[width=\mywidth]{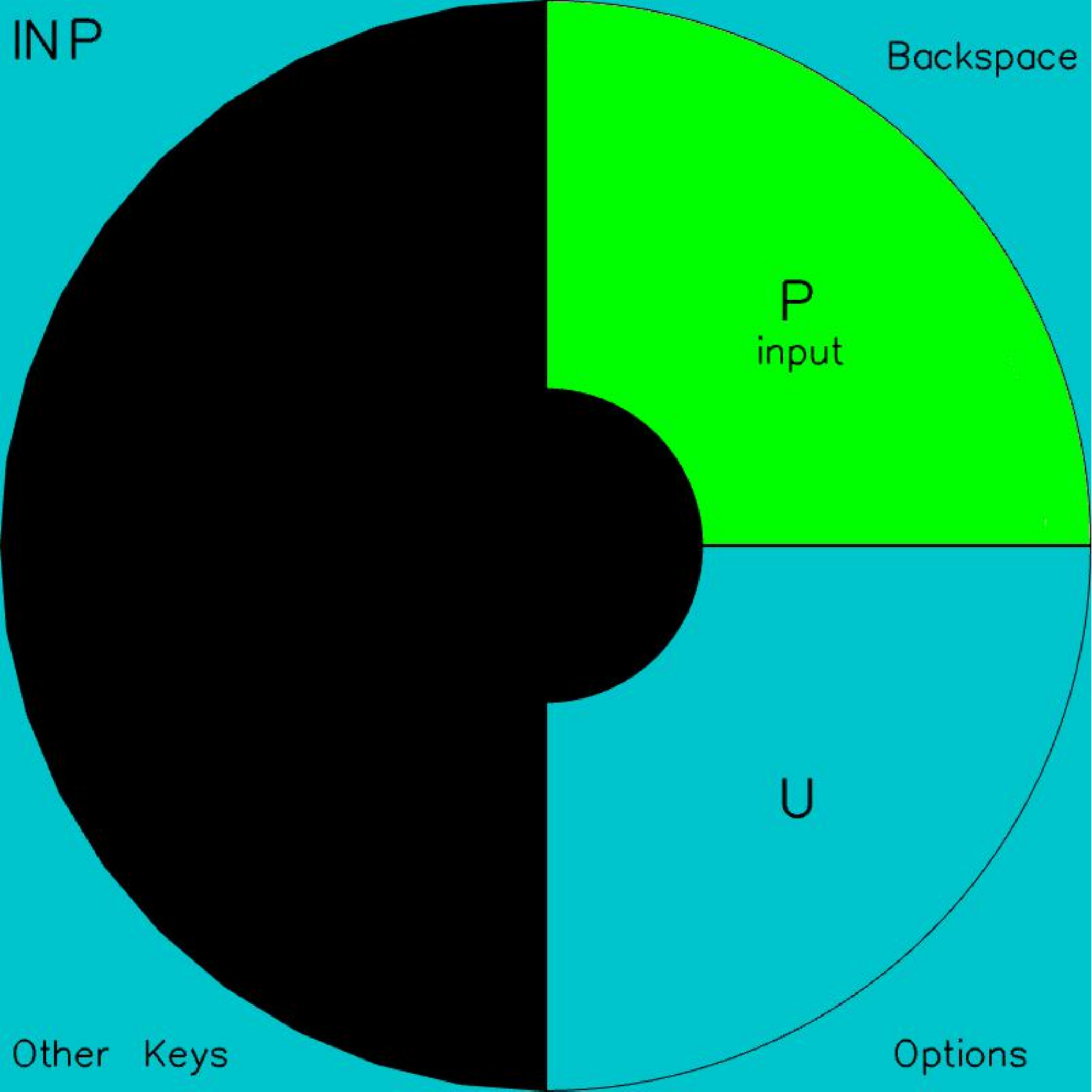}
	}
	\caption {The user has entered the word prefix ``\texttt{in}". 
		(a) The current focus is on the letter `\texttt{p}', and SliceType suggests the word ``\texttt{input}" for completion. 
		(b) The user is about to complete the selection of letter `\texttt{p}'. 
		(c) The user continues dwelling inside the letter `\texttt{p}' to select the prediction. 
		The second dwell period progress is illustrated in green. 
		(d) The user is about to complete the selection of the prediction, ``\texttt{input}". }
	\label{fig:SliceTypeUsage2}
\end{figure}

When the cursor is moved inside the boundaries of a key, it is highlighted in a light orange color.
This is illustrated in Figure~\ref{fig:SliceTypeUsage1}~(a), where the user has just moved the cursor inside the letter `\texttt{i}'.
To select this letter, the user has to stay inside the key for the entire dwell period.
The amount of dwell period that has passed since the cursor has moved inside the key is indicated by the dark orange color that continuously fills the key.
In Figure~\ref{fig:SliceTypeUsage1}~(b), about 50\% of the dwell period is up, and in Figure~\ref{fig:SliceTypeUsage1}~(c), about 90\% of the dwell period is up.
When the entire dwell period is up, the letter `\texttt{i}' becomes selected and appears on the top-left corner of the interface, as shown in Figure~\ref{fig:SliceTypeUsage1}~(d).
Also notice in Figure~\ref{fig:SliceTypeUsage1}~(d) that after the letter `\texttt{i}' is selected, some letters have been removed from the interface, and their space have been merged into adjacent keys. 
In Figure~\ref{fig:SliceTypeUsage2}, not all blank space is merged to the remaining keys, because we limited the merging to only nearby keys.
More aggressive merging changes the layout excessively, which is reported to be harmful~\cite{Gunawardana:2010}.

Figure~\ref{fig:SliceTypeUsage1} and Figure~\ref{fig:SliceTypeUsage2} demonstrate the details of SliceType's word suggestion and selection mechanism.
Assume that the user wants to type the word ``\texttt{input}".
As soon as the user moves the cursor inside the first letter of the word, `\texttt{i}', its color changes to light orange and the first suggested word ``\texttt{in}" appears inside the key of letter `\texttt{i}'.
Since the user desires to type ``\texttt{input}", rather than ``\texttt{in}", they complete the selection of letter `\texttt{i}' by dwelling inside its key for a single dwell period, and move on to the next letter, `\texttt{n}'.
SliceType continues suggesting the same word, ``\texttt{in}"; but this time, the suggested word is displayed inside the key of letter `\texttt{n}' (see Figure~\ref{fig:SliceTypeUsage1}~(e-f)).
After letter `\texttt{n}' is selected and appears on the top-left corner, the user now moves on to the next letter, `\texttt{p}'.
This time, SliceType suggests the word ``\texttt{input}", which is displayed inside the key of letter `\texttt{p}' as shown in Figure~\ref{fig:SliceTypeUsage2}~(a).
The user stays inside letter `\texttt{p}' for the entire dwell period to select it as shown in Figure~\ref{fig:SliceTypeUsage2}~(b), and when the selection is complete, letter `\texttt{p}' appears on the top-left corner, as shown in Figure~\ref{fig:SliceTypeUsage2}~(c).
Now, the user wants to select the suggested word ``\texttt{input}" displayed inside the key of letter `\texttt{p}', so that they do not have to type the rest of the desired word.
To select the suggested word, the user must continue dwelling inside letter `\texttt{p}' for an additional dwell period.
This is illustrated in Figure~\ref{fig:SliceTypeUsage2}~(c), where the dwell period progress is illustrated by the green color.
When this second dwell period is up, the suggested word ``\texttt{input}" is selected, and is sent to the system's character input stream.
SliceType interface returns back to its default state shown in Figure~\ref{fig:SliceTypeGUI} to display all characters again, and the user can start typing a new word.

Words with repeating letters are a common occurrence in English.
Consider the word being entered to be ``\texttt{winning}".
While the user is dwelling to enter the first `\texttt{n}', the corresponding prediction would be ``\texttt{window}".
After dwelling once on `\texttt{n}', continuing to double-dwell would result in the prediction to be entered, rather than a second `\texttt{n}'.
Then, the user needs to exit the key `\texttt{n}' after the first dwell towards an arbitrary location, and return back to the same key.
Following from that, the keyboard starts entering the second `\texttt{n}', and proposes ``\texttt{winning}" as the prediction.
Now, the user can double-dwell to enter the desired word.

SliceType does not have a dedicated area on its interface to display the transcribed text.
Instead, it sends the entered text to the system's character input system, which can then be read by the program that has the current system focus.
This way, the entered text can appear inside any program, e.g., word processor, text-to-speech program, web browser, etc.

%~~~~~~~~~~~~~~~~~~~~~~~~~~~~~~~~~~~~~~~~~~~~~~~~~~~~~~~~~~~~~~~~~~~~~~~~~~~~~~~~~~~~~~~~~~~~~~~~~~~~~~~~~~~~~~~~~~~~~~~~~~~~~~~~
\section{Fitts' Law Analysis}
\label{sec:fitts}

The Fitts' law predicts the time required to move a pointing device to a target~\cite{Fitts:1954}.
In this study, we used its Shannon variant, which was proposed by MacKenzie~\cite{MacKenzie:1992}.
The formulation that estimates the movement time is as follows.
Where $A$ represents the distance between the movement origin and the target center, $W$ represents the width of the target, and $a, b$ are model parameters:

\begin{equation}
MT = a + b \log_2\left(\frac{A}{W} + 1\right)
\end{equation}

To estimate the time of a specific movement, one must estimate $a$ and $b$.
Our objective is to quantify the relative performances expected with different keyboard settings.
Thus, we will use the form in which only the index of difficulty (ID) is estimated:

\begin{equation}
ID = \log_2\left(\frac{A}{W} + 1\right)
\label{eq:fitts}
\end{equation}

\begin{figure}
	\centering
	\includegraphics[width=0.5\columnwidth]{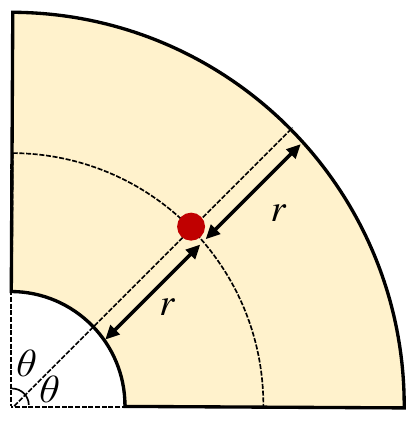}
	\caption {The target center (shown in red) is defined to be the intersection of the radial symmetry axis, and the concentric circle that is equidistant from the circular borders.}
	\label{fig:fl1}
\end{figure}

While ID does not indicate the time required for an action directly, we can say that an action with higher ID will take longer, because $a$ and $b$ are defined to be positive.
MacKenzie and Buxton propose a method for using the Fitts' law with 2D targets~\cite{MacKenzie:1992b}.
The $W'$ model proposed in this study replaces $W$ with the extent of the target along the line that passes through the movement origin and the target center.
To use this model, the target center has to be defined.
For the key shapes in SliceType, we used the intersection of the radial symmetry axis of the key and the concentric circle that is equidistant from the circular borders of the key (see Figure~\ref{fig:fl1}).
The same target definition is used for the two half circle keys in the center of the keyboard.
For these two keys, the inner circular borders are assumed to have a radius of $0$.

\begin{figure}
	\centering
	\includegraphics[width=0.8\columnwidth]{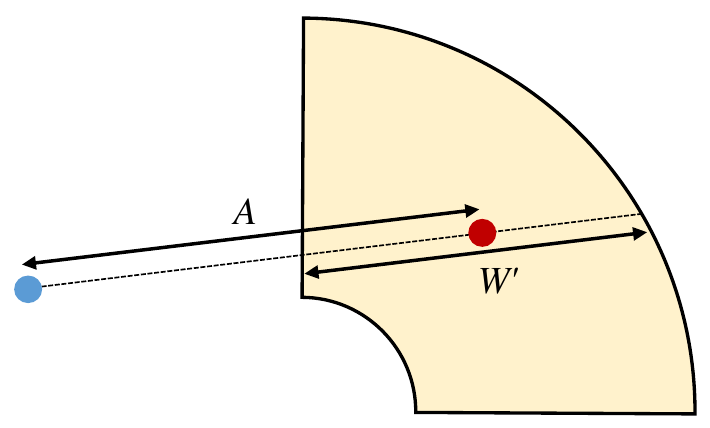}
	\caption {The defined width, $W'$, is the extent of the target along the line that passes through the movement origin (shown in blue) and the target center (shown in red).
		$A$ is the distance between the movement origin and the target.}
	\label{fig:fl2}
\end{figure}

After defining the target center, calculating $W'$ and $A$ is rather straightforward.
To find $W'$, a line passing through the movement origin and target center is drawn.
The distance between the two intersections of this line with the key gives the target width (see Figure~\ref{fig:fl2}).
$A$ is the distance between the movement origin and the target center.
These variables are used in Eq.~\ref{eq:fitts} to find the ID of each movement.

The pangram, ``The quick brown fox jumps over the lazy dog'', is used to calculate the Fitts' law ID scores.
The comparison is done between combinations of utilizing two features of the proposed soft keyboard.
The first feature is instantly completing the word using the proposed prediction.
The second feature is the merging of the keys to increase target width.

\subsection{Without Word Prediction and Key Merging}

In this condition, the whole sample text is written letter by letter, i.e., the predictions proposed by the soft keyboard are not used.
This increases the total distance to be traveled, which directly increases the Fitts' law cost.
Furthermore, the keys that do not propose a prediction are not removed, thus the free space is not used to enlarge the remaining keys.
Since the key widths are less than ideal, the Fitts' law cost is again increased.
In other words, this condition is expected to produce the worst results.
According to the methodology described earlier, the Fitts' law ID for the test pangram is calculated to be $49.02$.

\subsection{With Word Prediction and without Key Merging}
\label{subsec:wpnomerge}

Only word prediction is utilized in this condition. The user is assumed to use the true predictions as soon as they are proposed, yet the keys that do not propose a prediction are not removed.
This condition actually simulates the default usage scenario of the majority of soft keyboards.
The Fitts' law ID is calculated to be $40.40$.

\subsection{Without Word Prediction and with Key Merging}

Contrary to the previous condition, keys that do not produce a word prediction are merged to the remaining keys.
However, the user is assumed not to utilize the predictions, i.e., all letters are typed individually.
The calculated Fitts' law ID for this condition is $33.68$.

\subsection{With Word Prediction and Key Merging}

This condition assumes the normal usage of SliceType as discussed in Section~\ref{sec:slicetype}.
The predictions are utilized as soon as they are proposed and the keys that do not propose a prediction are removed to make room for the remaining keys.
This is the best case for user performance, thus the Fitts' law ID is expected to be the lowest.
Indeed, the calculated value is only $30.13$.

\subsection{Placement of Predictions}

In Section~\ref{sec:keylayout}, the importance of the method of delivery for prediction proposals was emphasized.
Placing all predictions in a single area of the layout has many downsides.
Instead, we have proposed to place each related prediction inside a key.
In this part, we will try to quantify the achieved user performance improvement using the Fitts' law.
Since key merging is irrelevant in this matter, we will use the Fitts' law ID result of the second condition (see Sec.~\ref{subsec:wpnomerge}).
When the user uses the predictions in the keys, the Fitts' law ID was calculated to be $40.40$. 

The alternative way of presenting the predictions was assumed to be placing them at the top-left corner.
Obviously, not all predictions will fit in that area, as we propose up to 26 predictions at a time.
However, we will assume that all predictions can fit in the top left area, and the user can read them as easily as they read them in the proposed method.
The user is assumed to check the predictions after typing each letter to see if the word to be entered is among the predictions.
This adds a substantial amount of extra gaze movement, which directly translates as a Fitts' law cost.
Using the same sample text, the Fitts' law ID score is calculated to be $109.37$.
We can see that simply moving the predictions inside the keys results in a $63.1\%$ decrease in the Fitts' law ID.

\subsection{Analysis of the Results}

\begin{table}
	\centering
	\renewcommand{\arraystretch}{1.4}
	\caption{Fitts' Law Index of Difficulty (ID) Results}
	{
			\begin{tabular}{ l | c | c }
				& Without Prediction & With Prediction \\ \hline
				Without Key Merging & $49.02$ & $40.40$ \\ \hline
				With Key Merging & $33.68$ & $30.13$ \\
			\end{tabular}
			\label{tab:FittsID}		
	}
\end{table}

The Fitts' law indices of performance (IDs) calculated under different conditions are presented in Table~\ref{tab:FittsID}.
As expected, adding the word prediction and key merging features resulted in a decrease in ID, which will translate to faster text entry.
The case in which both predictions are utilized and keys are merged performed best with an ID of $30.13$.
In contrast with this result, the condition in which predictions and key merging are disabled performed worst with an ID of $49.02$.
The difference cannot be interpreted as a $38.5\%$ speed up, because as we have discussed in the beginning of this section, ID is multiplied by a parameter ($b$) and the result is added with another parameter ($a$) to find the total time needed for the user to travel through the designated path.
However, we can assume that lower ID will correspond to faster typing.

The more interesting results are seen for the cases in which only one of the features is implemented.
Word prediction is commonly implemented by soft keyboards as it is obvious that typing less characters will result in faster text entry.
However, we see that the proposed key merging functionality will improve performance significantly more than using word predictions.
Only using word predictions results in an ID of $40.40$, while only using key merging results in an ID of $33.68$.
We can say that soft keyboard designers should invest more of their focus in effective allocation of area, as doing so may produce better results, even if the total number of actions to enter a piece of text would increase.

%~~~~~~~~~~~~~~~~~~~~~~~~~~~~~~~~~~~~~~~~~~~~~~~~~~~~~~~~~~~~~~~~~~~~~~~~~~~~~~~~~~~~~~~~~~~~~~~~~~~~~~~~~~~~~~~~~~~~~~~~~~~~~~~~
\section{Experiments}

In this Section, we present the results of our experiment where we compared SliceType with two other publicly available gaze typing keyboards, Dasher~\cite{Ward:2000} and GazeTalk~\cite{Hansen:2001}.
These keyboards were chosen to cover a wide variety of design choices.
SliceType and GazeTalk perform dwelling based selection, while Dasher employs continuous gestures.
Itoh et al. have compared GazeTalk and Dasher for typing in Japanese, and predicted Dasher to be better in the long-term~\cite{Itoh:2006}.
To our knowledge, this is the only other study in the literature that compared gaze typing soft keyboards directly.

\subsection{Experimental Design}
\label{sec:expdes}

All three keyboards are operated using a continuous input device.
To compare them in terms of gaze typing performance, we used an eye tracking system to translate the user's eye movements to cursor movements.
The eye tracker used in the experiment is Monocular Edge Analysis System from LC Technologies.
It has a sampling rate of 60~Hz, and a gaze position estimation accuracy of $0.45\degree$.
In the experiments, the eye tracker is used from a typical distance of 50~cm.
The calibration was repeated before the use of each keyboard.

The experiment was conducted with 37 undergraduate and graduate computer and electrical-electronics engineering students.
None of the participants have used an eye tracker before.
The participants were not asked to remove their glasses or contact lenses.
They were given written usage instructions for each keyboard prior to the experiment.
After reading the instructions, each participant was individually taught how the eye tracking system and each keyboard works for a total of thirty minutes.
After this brief training session, the participants were asked to gaze-type a different paragraph with each keyboard.
The paragraphs were chosen to represent the complexity of daily conversation.

Instead of having the participants read from the source material, we dictated it to them.
This was done to emulate typing as a means of communication.
The participants were asked to type as much of the text as possible in 5 minutes.
The participants were instructed to correct any errors they may make before the session, but they were not actively notified of their errors during the session.

Both the keyboard order and keyboard--paragraph matching were permuted to prevent any bias.
The keyboards were sized to cover the left-half of a 21" screen with $1920\times1080$ resolution.
The parameters for each keyboard were kept as their default values.
That is, the dwell period for both SliceType and GazeTalk were set to 1000 ms, and for Dasher, the speed was set to 0.8 with adaptive speed adjustment enabled.

After the experiment was finished, the participants were asked which keyboard they would prefer to use if operating a soft keyboard with an eye tracker was their only means of communication.
The participants put the keyboards in the order of preference.
The reasoning behind this question was that although a user may type slowly using a keyboard, they may had a better user experience with it, and vice versa.
The experiment and the survey were done blindly, meaning that the participants were not informed about which keyboard was being developed in this study.

\subsection{Results}

Let us begin by emphasizing again that the experiment were conducted with novice users.
Both the text entry rates and preferences are likely to change as the users gain proficiency.
Therefore, the results should be considered as an early prediction, rather than a conclusion.

\begin{figure}
	\centering
	\includegraphics[width=0.8\columnwidth]{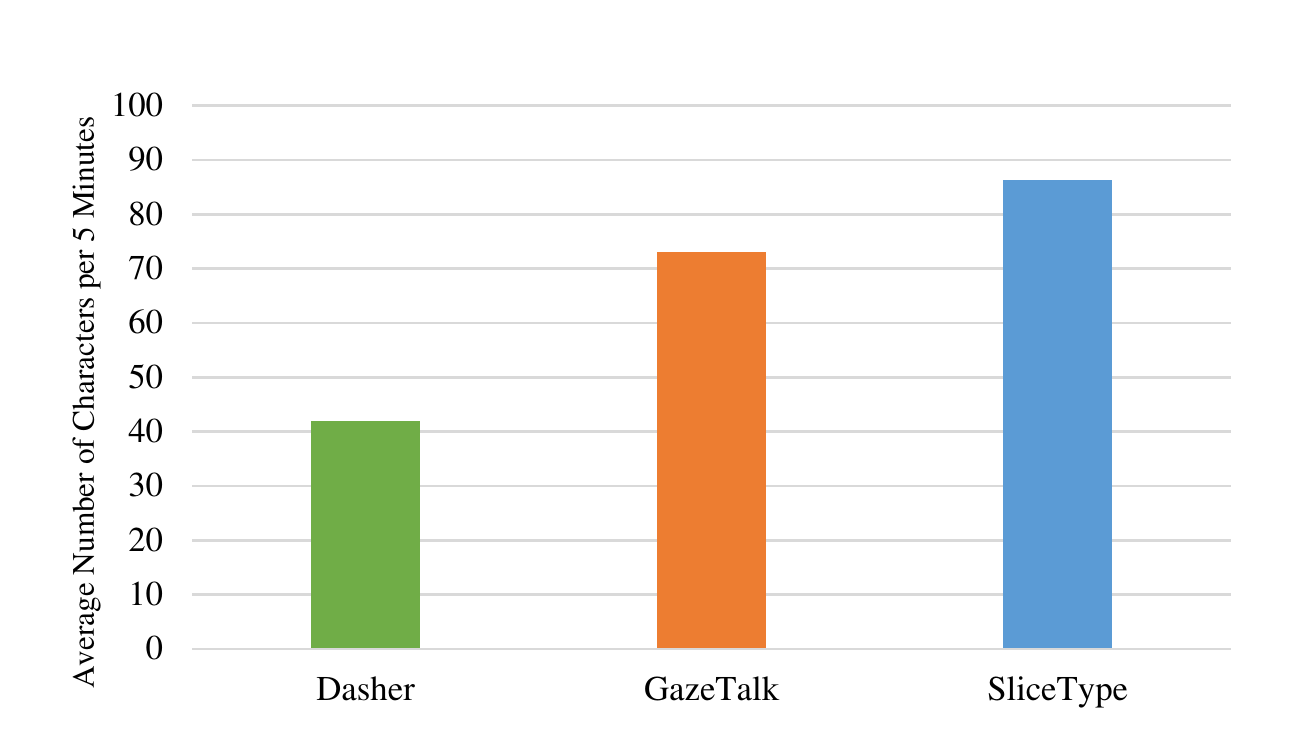}
	\caption {Average number of characters entered by the participants in 5 minutes.}
	\label{fig:Exp1}
\end{figure}

Figure~\ref{fig:Exp1} illustrates the average number of characters entered by all participants in 5 minutes.
Clearly, SliceType provides the highest text entry rate, while the participants performed poorly with Dasher.
We compared these results with the ones of our previous experiment~\cite{Topal:2012} in Table~\ref{tab:resTable}.
The results from Figure~\ref{fig:Exp1} are converted to words per minute (wpm) using the 5 characters per word standard.
It is important to note that the soft keyboard in~\cite{Topal:2012}, eOSK, was an early prototype of SliceType that did not include key merging.
Therefore, the mouse experiment result for SliceType is not very representative.

\begin{table}
	\centering
	\renewcommand{\arraystretch}{1.4}
	\caption{Text Entry Rates (Word per Minute)}
	{
			\begin{threeparttable}
				\begin{tabular}{l|c|c|c} 
					& Dasher & GazeTalk & SliceType \\ \hline
					Mouse Experiment~\cite{Topal:2012} & 3.50 & 3.07 & 5.42* \\ \hline	
					Eye Tracker Experiment & 1.68 & 2.93 & 3.45 \\ 	
				\end{tabular}
				\begin{tablenotes}
					\item[*] Without key merging.
				\end{tablenotes}
			\end{threeparttable}
		\label{tab:resTable}
	}
\end{table}

As seen in Table~\ref{tab:resTable}, SliceType has the highest text entry rate in both experiments, but the ranking between Dasher and GazeTalk is reversed.
The difference between the results obtained in this experiment and the results from our previous experiment in~\cite{Topal:2012} has an obvious explanation.
The experiment in~\cite{Topal:2012} was conducted using a mouse as the input device, while the experiment presented in this paper is conducted with an eye tracker.
Table~\ref{tab:resTable} shows that the usage of the eye tracker nearly halved the number of words entered using Dasher and SliceType, while it did not affect GazeTalk's performance much.
Both Dasher and SliceType require high fidelity input, yet GazeTalk's larger keys allow the use of a less accurate input device.
Even though SliceType's performance deteriorated when used with an eye tracking system, it still performs the best among the three keyboards.

\begin{table}
	\centering
	\renewcommand{\arraystretch}{1.4}
	\caption{Post-hoc Comparisons with Bonferroni Correction Following One-way ANOVA ($p<0.001$) to Compare the Effect of Keyboard on Text Entry Rate in 5 Minutes (in Number of Characters)}
	{\hspace{-16mm}
		\parbox{.45\linewidth}{
			\centering
			\begin{tabular}{c|c}
				
				& Mean \\ \hline
				Dasher & 42 \\ 
				GazeTalk & 73.14 \\ 
				SliceType & 86.30
			\end{tabular}
		}
		\parbox{.45\linewidth}{
			\centering
			\begin{tabular}{c|c}
				& p value \\ \hline
				SliceType--Dasher & $<0.001$* \\
				SliceType--GazeTalk & 0.193	\\
				GazeTalk--Dasher & $<0.001$*
			\end{tabular}
		}
		\label{tab:testEntryRate}
	}
\end{table}

To compare the mean text entry rates of keyboards, we conducted a one-way ANOVA, which indicated significant entry rate variance with respect to the keyboard type ($F(2,108)=20.871$,~$p<0.001$ ).
To analyze the results further, we applied post-hoc comparisons with Bonferroni correction~(see Table~\ref{tab:testEntryRate}).
It can be seen that there is a very significant difference between the text entry rates achieved with Dasher and SliceType.
The difference between GazeTalk and SliceType is less pronounced ($p\nless0.05$).
As an additional result of our experiment, we have observed that the text entry rate was significantly higher with GazeTalk, compared to Dasher.

\begin{figure*}
	\centering
	\includegraphics[width=\textwidth]{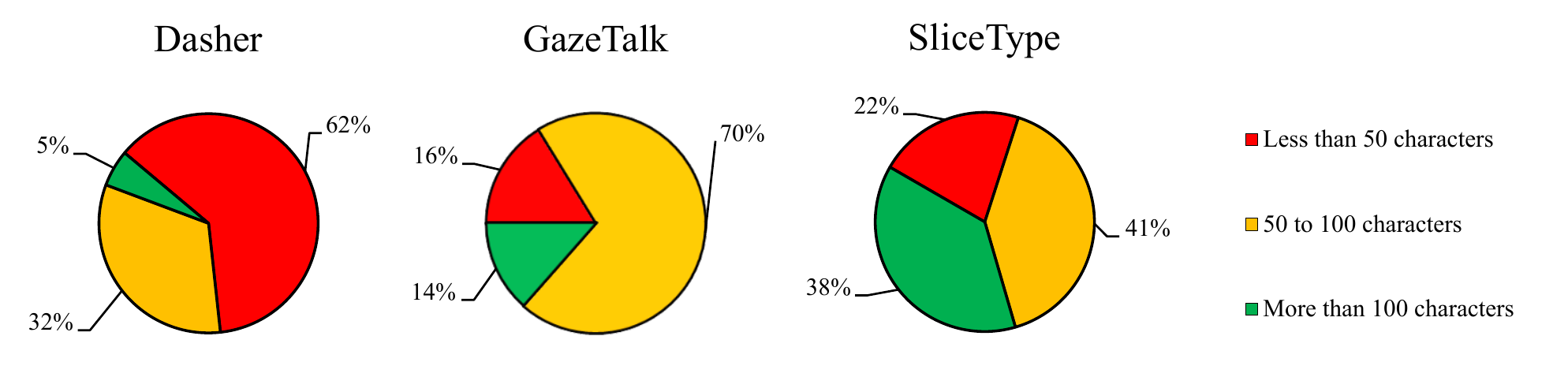}
	\caption {Percentage of the participants who typed $<$50, $50$--$100$, or $>$100 characters with different keyboards during the 5 minute text entry session.}
	\label{fig:PieCharts}
\end{figure*}

In Figure~\ref{fig:PieCharts}, the participants were grouped based on their character entry rate with each keyboard.
That is, the percentages of the participants who typed less than 50 characters, in between 50 and 100 characters and more than 100 characters using each keyboard are illustrated.
It is clear from the figure that the majority of the users were not able to use Dasher efficiently.
This shows the difficult learning curve of Dasher.
With GazeTalk, most of the users were able to type at an average speed.
Only a few users were able to reach very high typing speeds of 100 or more characters.
With SliceType, many participants were able to reach an average typing speed of 50 to 100 characters, while as many participants were able to reach high typing speeds of more than 100 characters.
This shows that SliceType is easy to learn and use, and the users can speed up even more with a little practice.
Only a small percentage of the participants were not able to use SliceType efficiently.
If we examine the average entry rate of these participants across keyboards, we can see that they are all below average.
This implies that the participants who underperformed with SliceType may have suffered from a more general problem.
This can include variations in eye tracker performance based on the individual, participants' level of motivation and proficiency with computers.

\begin{figure*}
	\centering
	\includegraphics[width=\textwidth]{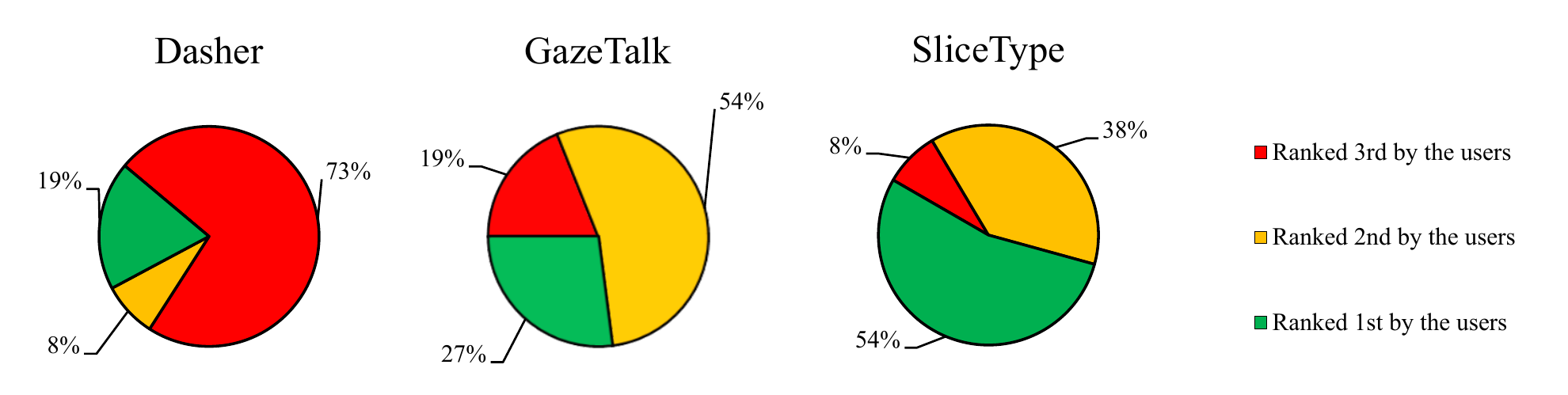}
	\caption {The users were asked to rank the keyboards in their order of preference for future use.
		Lower rankings indicate higher preference.
		The results are presented as percentages.}
	\label{fig:Exp2}
\end{figure*}

Finally, the participants were asked to rate the keyboards in terms of their order of preference.
The results are illustrated in Figure~\ref{fig:Exp2}.
Recall that both the experiments and the survey were conducted blindly to prevent any bias towards a keyboard.
The participants were also not informed of their text entry rates with the keyboards before filling out the survey.
Intuitively, we can expect the users to prefer the keyboards that they can type faster with.
The results of the survey can be summarized as follows:

\begin{enumerate}
	\item 46\% of the users ranked the keyboards in the same order with their typing speed.
	\item 59\% of the participants ranked the keyboard that they typed fastest with as the best.
	\item 65\% of the participants ranked the keyboard that they typed slowest with as the worst. 
\end{enumerate}

\begin{figure}
	\centering
	\includegraphics[width=0.8\columnwidth]{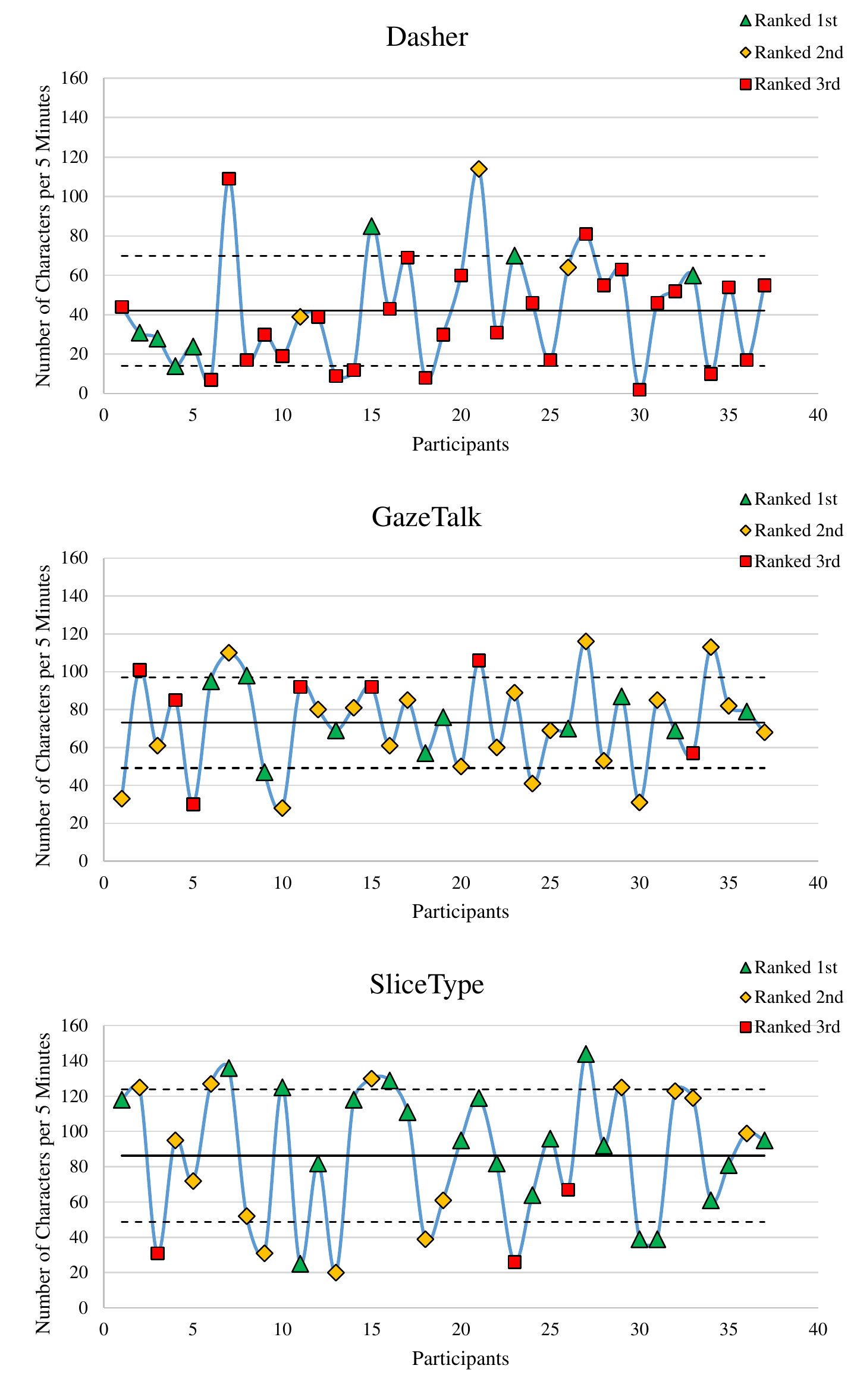}
	\caption {Number of characters entered in 5 minutes by each participant, along with the participant's ranking of the keyboard. 
		``Ranked 1st" indicates that the keyboard is the most preferred and ``Ranked 3rd" indicates that the keyboard is the least preferred by the participant. 
		The solid black line indicates the mean, while the dashed lines indicate $\pm1$ standard deviation.}
	\label{fig:Exp3}
\end{figure}

While it is obvious that there is a correlation between typing speed and preference, it may not be as strong as it is assumed to be.
41\% percent of the users do not prefer the keyboard that they type fastest with.
This may indicate that focusing solely on the text entry speed while designing a soft keyboard is rather questionable.
68\% of the participants type fastest with SliceType, while only 54\% of the participants prefer SliceType as the best keyboard.
This shows that any design improvements to SliceType should be on usability, rather than text entry speed.
Participants' individual performances and preference rankings are shown in Figure~\ref{fig:Exp3}.

\begin{table}
	\centering
	\renewcommand{\arraystretch}{1.5}
	\caption{Post-hoc Comparisons with Bonferroni Correction Following Aligned-rank Transform and One-way ANOVA ($p<0.001$) to Compare the Effect of Keyboard on User Preference}\medskip
	
	\parbox{.45\linewidth}{
			\centering
			\begin{tabular}{c|c}
				& \pbox{3cm}{\centering Mean aligned ranks} \\ \hline
				Dasher & 36 \\ 
				GazeTalk & 59 \\ 
				SliceType & 73
			\end{tabular}
	}
	\parbox{.53\linewidth}{
			\centering
			\begin{tabular}{c|c}
				& p value \\ \hline
				SliceType--Dasher & $<0.001$* \\
				SliceType--GazeTalk & $0.074$	\\
				GazeTalk--Dasher & $0.001$*
			\end{tabular}
	}
	\label{tab:testRanking}
\end{table}

We applied aligned-rank transform to the ordinal survey data, and conducted a one-way ANOVA test, similar to the test done for text entry rate.
The results indicated very significant variance in preference with respect to the keyboard ($F(2,108)=18.476$,~$p<0.001$).
We followed up with post-hoc comparisons with Bonferroni Correction.
On Table~\ref{tab:testRanking}, it can be seen that in average, SliceType is ranked to be the best among the participants.
The preference of SliceType over Dasher is very significant, while the preference over GazeTalk is not strictly significant ($p\nless0.05$).
Similar to the character entry rates, we can observe that GazeTalk is significantly better preferred over Dasher.

\begin{table}
	\centering
	\renewcommand{\arraystretch}{1.4}
	\caption{One-way ANOVA ($p=0.094$) to Compare the Effect of Keyboard on the Number of Errors}
	\centering
	\begin{tabular}{c|c}
		& Mean \\ \hline
		Dasher & 1.027 \\ 
		GazeTalk & 0.541 \\ 
		SliceType & 0.676
	\end{tabular}
	\label{tab:testError}
\end{table}

In Section~\ref{sec:expdes}, we have mentioned that we have instructed the participants to correct their errors beforehand.
However, the entries by some participants included errors due to oversights, time limitation and a simple failure to operate the deletion mechanism of the keyboard.
To test if the keyboard type had any effect on the number of errors, we first need to define what is to be considered an error.
We defined consecutive incorrect characters to be a single error.
For example, entering ``ktboard" instead of ``keyboard" counts as a single error.
According to this definition, we applied one-way ANOVA, and observed that the number of errors did not change with different keyboards significantly ($F(2,108)=2.417$,~$p=0.094$, see Table~\ref{tab:testError}).

%~~~~~~~~~~~~~~~~~~~~~~~~~~~~~~~~~~~~~~~~~~~~~~~~~~~~~~~~~~~~~~~~~~~~~~~~~~~~~~~~~~~~~~~~~~~~~~~~~~~~~~~~~~~~~~~~~~~~~~~~~~~~~~~~
\section{Conclusion}

The main objective while designing a soft keyboard is to allow fast text entry with a comfortable user experience.
In this paper, we investigated the specific case of gaze typing keyboards.
Increasing the target sizes is the most effective way of improving a dwelling keyboard.
That is because large keys are fast to navigate towards, and easy to dwell on.
However, statically displaying a few large keys is not ideal.
Instead, we focused on dynamically enlarging the probable target keys to use the interface area efficiently.
Given a fixed sized interface, the only way to enlarge a key is to shrink others.
We proposed a more extreme approach, namely removing the keys that are not likely to be used.
Our analysis based on the Fitts' law showed that this key merging approach improves text entry rate even more than word completion.

Word completion is a fundamental tool, used in nearly all contemporary soft keyboards.
While the language model used to produce predictions is important, a more subtle factor is how the predictions are communicated to the user.
The common way is to present predictions in a dedicated area, grouped together.
The user has to look over to the area to read the predictions, and going through a long list takes time.
We propose to display a prediction at each key, so that the user will already be looking at the prediction while dwelling to select a character.
The Fitts' law analysis shows that this modification has a critical impact.
Moreover, since these predictions are displayed individually, it is fast, even involuntary, to read them.
A downside of this approach is that the keys not only have to be large enough to contain characters, but also words.
However, the key merging function enlarges the majority of the target keys, thus is complementary as a solution to this problem.

We experimentally compared the keyboard designed using these principles, SliceType, with two other gaze typing keyboards.
The results showed that novice users typed faster with SliceType, and preferred it over others for daily communication.
This validates that the design principles we have proposed are in accordance with our objectives.
An interesting result is that 68\% of the participants typed fastest with SliceType, while only 54\% preferred it over the other two.
This indicates that further study on the subject should be more focused on user comfort than text entry rate.

\small
\bibliographystyle{IEEEtran}
\bibliography{slicetype-bibfile}

\end{document}